\documentclass[12pt]{article}
\usepackage[centertags]{amsmath}
\usepackage{amssymb,amsfonts}
\usepackage[dvips]{graphicx}
\usepackage{epsfig}

\makeatletter \@addtoreset{equation}{section} \makeatother

\newcommand{\tr}{\operatorname{tr}}

\begin{document}

\newcommand{\sk}[1]{\textbf{sk: {#1}}}

\begin{titlepage}

\begin{center}
\vspace{-2cm}

\hfill {Imperial/TP/08/SK/01}\\
\hfill  {\tt arXiv:0803.4183} \\
         [12mm]

{\Huge Comments on 1/16 BPS Quantum States and Classical Configurations}
\vspace{8mm}

{\large Lars Grant$^{a,b}$, Pietro A. Grassi$^{c}$, Seok Kim$^{d,e}$ and Shiraz Minwalla$^{a}$} \\
\vspace{3mm}
{$^a$\small \it Department of Theoretical Physics, \\
\small \it Tata Institute of Fundamental Research, \\} {\small \it
Homi Bhabha Road, Colaba, Mumbai - 400005, India.\\}

\vspace{1mm} {$^b$\small \it Jefferson Lab, Department of
Physics,\\} {\small \it Harvard University, \\} {\small \it
Cambridge, MA 02138, USA.\\}

\vspace{1mm} {$^c$\small \it Department of Sciences and Advanced Technologies \& INFN,\\} 
{\small \it University of Eastern Piedmont, \\} {\small \it
Alessandria, I-15100, Italy.\\}

\vspace{1mm} {$^d$\small \it Theoretical Physics Group, Blackett
Laboratory,\\} {\small \it Imperial College, London SW7 2AZ, U.K.\\}

\vspace{1mm} {$^e$\small \it Institute for Mathematical Sciences,\\}
{\small \it Imperial College, London SW7 2PE, U.K.\\}

\vspace{1mm} {\footnotesize Email: \ lars.grant@mail.mcgill.ca, \ pgrassi@cern.ch, \
s.kim@imperial.ac.uk, \ minwalla@theory.tifr.res.in}\\
\end{center}

\vspace{-3mm}

\begin{abstract}

We formulate the problem of counting 1/16 BPS states of ${\cal N}=4$
Yang Mills theory as the enumeration of the local cohomology of an
operator acting on holomorphic fields on $C^2$. We study aspects of
the enumeration of this cohomology at finite $N$, especially for
operators constructed only out of products of covariant derivatives
of scalar fields, and compare our results to the states obtained
from the quantization of giant gravitons and dual giants. We
physically interpret the holomorphic fields that enter our
conditions for supersymmetry semi-classically by deriving a set of
Bogomolnyi equations for 1/16-BPS bosonic field configurations in
$\mathcal{N}=4$ Yang Mills theory on $R^4$ with reality properties and boundary
conditions appropriate to radial quantization. An arbitrary solution
to these equations in the free theory is parameterized by
holomorphic data on $C^2$ and lifts to a nearby solution of the
interacting Bogomolnyi equations only when the constraints
equivalent to $Q$ cohomology  are obeyed.

\end{abstract}

\end{titlepage}


%


\tableofcontents

\section{Introduction}

As ${\cal N}=4$ Yang Mills theory is dual to IIB string theory on
$AdS_5 \times S^5$
\cite{Maldacena:1997re,Gubser:1998bc,Witten:1998qj,Aharony:1999ti},
exact results in this theory hold the promise of important lessons
about gravitational dynamics. A quantity that appears physically
interesting as well as potentially calculable is the exact finite $N$
partition function over supersymmetric states in this Yang Mills
theory quantized on $S^3$. In the bulk such a partition function
would characterize an interacting collection of supersymmetric
gravitons, giant gravitons
\cite{McGreevy:2000cw,Grisaru:2000zn,Hashimoto:2000zp} and black
holes at successively higher energies, and should contain a wealth
of information about the structure and dynamics of these objects.
This BPS partition function may also turn out to have mathematical
interest.

The partition function over states that preserve at least one eighth
of the supersymmetries of Yang Mills theory has already been
determined \cite{Kinney:2005ej}. Related issues are discussed in
\cite{Aharony,Berenstein:2005aa}. Moreover the most general index
over all supersymmetric states in this theory has also been
determined in terms of an integral over a single unitary matrix (at
finite $N$) \cite{Kinney:2005ej}. However the full finite $N$
partition function over all supersymmetric states - which includes
contributions from 1/16 BPS states - has yet to be found. As IIB
supergravity on $AdS_5 \times S^5$ hosts no regular 1/8 BPS black
holes but (at least) a four parameter class of 1/16 supersymmetric
black hole solutions \cite{Kunduri:2006ek} (see
\cite{Gutowski:2004ez,Gutowski:2004yv,Chong:2005da,Chong:2005hr}
also), the full supersymmetric partition function is likely to be
qualitatively richer than those that have been determined to date.

In this paper we report modest progress towards characterizing the
1/16 BPS partition function of ${\cal N}=4$ Yang Mills theory, which
we now proceed to describe.

Our paper is based on the following conjecture. We conjecture that
the supersymmetric spectrum of ${\cal N}=4$ Yang Mills theory on
$S^3$ is exactly given by the spectrum of 1/16 BPS states of the finite
$N$ one loop Beisert Hamiltonian of ${\cal N}=4$ Yang Mills
\cite{Beisert:2003jj,Beisert:2004ry}. This conjecture (which is
implicit in \cite{Kinney:2005ej} and \cite{Berkooz:2006wc}
and was explicitly stated in \cite{Minwalla:strings2006}) is as yet unproved but is consistent with all known data about the
supersymmetric spectrum of Yang Mills theory, including
strong-coupling countings
\cite{Kinney:2005ej,Biswas:2006tj,Mandal:2006tk}. Via the state
operator map, it is equivalent to the statement that the 1/16 BPS
spectrum of ${\cal N}=4$ Yang Mills is given by the nonlinear
classical cohomology of a particular supercharge in the
supersymmetry algebra (see section \ref{local}).

In section \ref{cohomologyapproach} of this paper we describe the classical action of this
special supercharge on `letters' of Yang Mills theory at the origin
of $R^4$. These `letters' are simply all covariant derivatives of
the basic gauge invariant fields -  scalars, field strengths and
fermions - of ${\cal N}=4$ Yang Mills theory (subject to the
operator equations of motion). The structure of the action of a
supercharge, $Q$, on
any letter $A$ of Yang Mills takes the form $Q A= [B,C]$ where $B$ and $C$ are
other letters, and the commutator involves matrix multiplication. It
follows immediately from this structure that $Q$ acts on products of
traces `one at a time'. This allows us to argue that the cohomology
of $Q$ at energies smaller than $N$ (when products of  traces form
an unambiguous basis of gauge invariant operators) is simply given
by the Bose Fermi multiparticling of the single trace $Q$
cohomology.

This result, which follows immediately from general considerations,
already makes an interesting prediction for bulk dynamics. Recall
that single trace operators are dual to bulk gravitons. Consider a
collection of bulk gravitons, each of which is individually
annhihilated by a particular supercharge. Our result implies that
the collection in question is also annihilated by the same
supercharge (i.e. that $1/N$ interaction effects do not lift the
energy of this collection above the sum of energies of its
consitutents) provided that the sum of energies of the gravitons is
smaller than $N$. We do not know how to derive this result from the
bulk, but it follows immediately as a consequence of our conjecture
about $Q$ cohomology. As the set of 1/16-BPS single trace Yang Mills
operators is already known (it consists of descendants of chiral
primaries), our argument lends support to the 1/16 BPS partition
function of Yang Mills theories at energies smaller than $N$
conjectured in \cite{Kinney:2005ej} (see also \cite{Janik:2007pm}).

In section \ref{cohomologyapproach} we also demonstrate that $Q$ simplifies
when acting on a local field which may mathematically be thought of as a
generating function for covariant derivatives of the letters of Yang Mills
theory. We will find a physical
interpretation of this generating function in section \ref{classicalconfigs}.

Having listed the action of the $Q$ operator, it is natural to
attempt to ennumerate its cohomology. We have achieved very modest
progress in this direction. Our concrete results pertain to special
subsectors of Yang Mills theory. First consider a subsector of
operators built out of any number of supersymmetric covariant
derivatives of a single supersymmetric scalar. It turns out to be
easy to evaluate the partition function over all operators in $Q$
cohomology which have at least one representative in this sector.
Our result in fact agrees with the partition function previously
obtained by Mandal and Suryanarayana by quantizing a related class of giant
gravitons \cite{Mandal:2006tk}. We point out that the naive
quantization of the relevant dual giant gravitons does not reproduce
the same result, and provide a physical explanation for this
discrepancy.

We then turn to a consideration of operators built out of arbitrary
numbers of derivatives of any of the three scalar fields of Yang
Mills theory. Already the cohomology ennumeration problem in this
subsector appears to be a complicated combinatorial problem that we
have not been able to solve in general, except for a simple case
(see section 3.2) in which we use the combinatorics of syzygy
\cite{Benvenuti:2006qr,Feng:2007ur} to find the exact partition
function. However we do present a rigorous upper bound on the growth
with charge of this partition function and demonstrate that this
growth is parametrically too slow (in the parameter $N$) to account
for black hole entropy. Consequently, black hole entropy is presumably
dominated by operators that are not Q equivalent to operators constructed
out of scalars. We leave the study of such operators to future work.

Finally, in sections \ref{five} and \ref{classicalconfigs} we attempt to obtain a physical interpretation
of the local generating fields that naturally appear in our description of
the action of $Q$ on
$1/16$ BPS letters. For this purpose it is useful to take a step backwards
from the quantum problem we have studied so far, and study the manifold of
classical 1/16 BPS configurations of $\mathcal{N}=4$ Yang Mills theory on
$S^3\times\mathbb{R}$ (or equivalently on $R^4$ with appropriate reality conditions) focusing on the bosonic sector. We will now discuss
this in some detail.

The bosonic subgroup of the $\mathcal{N}=4$ superconformal algebra
is $SO(4,2) \times SO(6)$. A basis for the Cartan subalgebra of this
algebra is given by the energy $E$ of states (the quantum number
under the timelike $SO(2)$ factor of $SO(4,2)$), the half integer
$J$ and $\bar{J}$ values of $SU(2)\times SU(2) \sim SO(4) \in
SO(4,2)$ and $H_1, H_2, H_3$, the generators of rotations in
orthogonal two planes in $SO(6)$.\footnote{We shall also use $J_1$,
$J_2$ rotating two orthogonal 2-planes, $J=\frac{J_1+J_2}{2}$ and
$\bar{J}=\frac{J_1-J_2}{2}$.} Let $Q$ denote the supersymmetry
operator of ${\cal N}=4$ Yang Mills, whose Cartan charges are given
by $E={1\over 2}$, $J=-{1\over 2}$, $\bar{J}=0$,
$H_1=H_2=H_3={1\over 2}$. It follows immediately from the
superconformal algebra that states that are annihilated by both $Q$
and its hermitian conjugate obey the BPS bound
$\Delta\equiv E-2J-H_1-H_2-H_3=0$. We prove a classical version of this
bound; in particular we demonstrate that the classical Noether
charge corresponding to $\Delta$ is equal to a sum of squares, and
so is positive definite. Classical bosonic configurations with
$\Delta=0$ obey a set of first order Bogomolnyi equations obtained by
setting each of these squares to zero.

While the first order supersymmetric equations are explicit, they
are nonlinear and we do not know how to solve them in general.
However these equations may be solved very simply when
$g_{YM}$ is set to zero, when the equations linearize. It turns out
that the supersymmetric solutions of the free theory are parameterized
by holomorphic data on the base $C^2=R^4$. The holomorphic data that
parameterizes free solutions may, infact, be identified with the
holomorphic generating functions on which the action of $Q$ was conveniently
defined, as described above.

It is then
natural to ask whether each supersymmetric solution at $g_{YM}=0$
admits a small perturbation to a supersymmetric solution at
infinitesimal $g_{YM}$. This question may be addressed
perturbatively and the answer is no. It turns out that
supersymmetric solutions of the free theory must satisfy an infinite
class of integrability constraints (the first of which is the
integrated Gauss Law) in order that they may be perturbed to
supersymmetric solutions at infinitesimal coupling.

Now the quantization of all supersymmetric states of free Yang Mills theory
simply yields the Fock Space of supersymmetric Yang Mills `letters'.
Suitably interpreting the constraint equations above as quantum constraints
that must be additionally imposed on Hilbert space, we recover a description
of the supersymmetric Hilbert space that reduces to the description of
$Q$ cohomology described in the first part of this paper, but this time with
a physical interpretation for the `generating function' fields as physical
Yang Mills fields propagating on physical $R^4$ but restricted to the
supersymmetric sector.

The plan of the rest of this paper is as follows. In section \ref{cohomologyapproach} we
state our cohomology problem, discuss the superposition of BPS
states with low energy, review the classical cohomology problem and
reformulate it using fields in terms of which the action of $Q$ is local.
We show in particular that the constraints of $Q$-cohomology for operators
constructed purely from bosonic letters
reduce to holomorphic, local gauge invariance and symmetrization
of these fields. In section \ref{partfunctions}, we consider counting the states of the
classical cohomology which are made of scalar letters. We obtain exact
partition functions in the single scalar sector and in the two
scalar sector with $U(2)$ gauge group, and also present an upper
bound partition function for the general scalar cohomology. Using
these results, we discuss the giant graviton interpretations and
also show that the degeneracy of these cohomologies grows too slowly to
form a black hole. In section \ref{five} we present a set of bosonic
classical $\frac{1}{16}$-BPS equations for $\mathcal{N}=4$ Yang-Mills
theory. In section \ref{classicalconfigs} we focus on the solutions of these equations in
weakly-coupled Yang-Mills theory. We show that the
solutions of free theory should obey infinitely many constraints to
be lifted to nearby solutions in the interacting theory. In section
\ref{quantization}, we attempt to radially quantize the system. Imposing a natural
quantization prescription in the weakly-coupled theory, we obtain a
final condition which is equivalent to that of the $Q$-cohomology.
We also obtain an interpretation of the local fields as physical
Yang-Mills fields on $\mathbb{C}^2$. Appendix \ref{transformations} reviews the
supersymmetry transformations. Appendix \ref{checks} provides an independent
check for our $SU(2)$ two-scalar partition function in sectors with
a few derivatives. Appendices \ref{D},\ref{coordapp},\ref{exactsolutions},\ref{G} treat the details of sections
\ref{five} and \ref{classicalconfigs} as well as presenting some exact solutions of our BPS
equation.

\section{The 1/16 BPS Cohomology}
\label{cohomologyapproach}

$\mathcal{N}=4$ Super-Yang-Mills on $S^3\times\mathbb{R}$ has a $PSU(2,2|4)$ group of
global symmetries. The fermionic generators of this algebra are 16
supersymmetries, $Q^i_{\alpha},\bar{Q}_{i\dot{\alpha}}$ and 16
super conformal generators $S_i^\alpha,\bar{S}^{i\dot{\alpha}}$.
These generators transform
in the $SU(2)\times SU(2)\times SU(4)_R$ bosonic subgroup and an
upper $SU(4)$ index $i=1\ldots4$ indicates a fundamental
representation, while lower indices are antifundamental. With radial
quantization, these generators satisfy the relations:
\begin{equation}\begin{split}
S_i^\alpha=(Q^i_\alpha)^\dagger\qquad
\bar{S}^{i\dot{\alpha}}=(\bar{Q}_{i\dot{\alpha}})^\dagger\\
\bar{Q}_{i\dot{\alpha}}=(Q^i_\alpha)^*\qquad
\bar{S}^{i\dot{\alpha}}=(S_i^\alpha)^*
\end{split}\end{equation}
where $*$ denotes complex conjugation and $\dagger$ denotes
hermitian conjugation.

We will be interested in states of the quantum theory which are
annihilated by the minimum number of supercharges.  These are $1/16$
BPS states, which are annihilated by only a single supersymmetry,
say $Q^4_-$ (written as $Q$ in the introduction),
and its hermitian conjugate $S^-_4$. That is, we are
interested in states that satisfy
$Q^4_-|\psi\rangle=S^-_4|\psi\rangle=0$. It will be convenient for
us to adopt a slightly formal description of these $1/16$ BPS
states: If we formally regard $Q^4_-$ as an exterior derivative $d$
and $S^-_4$ as its Hermitian conjugate $d*$, then $\{Q^4_-,S^-_4\}$
corresponds to the Laplacian $\Delta=d*d+dd*$. Standard arguments
show that states with $\Delta=0$, which are harmonic forms, are in
one-to-one correspondence with states in the cohomology of $d$.
Analogous arguments, formulated in terms of $Q^4_-,S^-_4$ show that
states which satisfy $\{Q^4_-,S^-_4\}|\Psi\rangle=0$ are in
one-to-one correspondence with states in the cohomology of $Q^4_-$.

Therefore, from now on, we will consider the set of $1/16$ BPS
states to be either all states that are annihilated by both $Q^4_-$
and $S_4^-$, or all states that are $Q^4_-$ closed but not $Q^4_-$
exact. From the point of view of calculating a partition function
over the $1/16$ BPS states, the two formulations are equivalent.

\subsection{Cohomology at Energies less than $N$}

We will first consider the BPS cohomology of $\mathcal{N}=4$ at energies less
than $N$ where we can characterize it fully.

Consider any basis for single trace operators in Yang Mills theory.
The key observation here is that the action of $Q$ on Yang Mills fields
takes the form $Q A= [B, C]$ where each of $A, B, C$ are adjoint Yang |
Mills letters (see Appendix \ref{transformations}). It follows that the action of $Q$
on a single trace
operator once again returns a single trace operator \footnote{At high
energies it may be possible to use trace identities to rewrite this operator
as a polynomial in multi traces, but one may always choose not to do so.
At any event we eventually focus, in this section, on energies less than $N$
where such a rewriting is impossible.}. Consequently it is possible to choose
a graded basis in the space of all single trace operators.
Let $\{ \gamma_i \}$ represent any  basis of single trace operators such that
$[Q, \gamma_i] \neq 0$, and let $\{ \gamma_i, Q \gamma_i, \alpha_i \}$
represent a basis of all operators in the theory. As $\alpha_i$ are linearly
independent of all states that are not $Q$ closed ($\{\gamma_i\}$) and
all states that are $Q$ exact $\{ Q \gamma_i \}$, the set $\{ \alpha_i \}$
is a basis for $Q$ cohomology in the single trace sector. We denote the
linear space spanned by $\{ \alpha_i \}$ by ${\cal H}_{SUSY}^{ST}$.

We will now argue that at energies (or scaling dimensions) less than $N$,
the full BPS cohomology of Yang Mills theory is given by
${\cal F}({\cal H}_{SUSY}^{ST})$. This result follows immediately from the
observations that
\begin{enumerate}
\item The fock space of single trace operators constitutes basis
for the space of all gauge invariant operators at scaling dimensions less
than $N$.
\item
The action of $Q$ preserves
the number of traces, and moreover $Q$ acts in a trace by trace manner;
for example
\begin{equation} \label{produ}
[Q,AB]=[Q,A]B\pm A[Q,B]=0\ ,
\end{equation}
where $A$ and $B$ are single trace operators.
\item The mathematical result that the fock space cohomology
of an operator $Q$ is the fock space of its cohomology (see for instance the discussion surrounding
equation 12.4.23 of \cite{Green:1987mn}).
\end{enumerate}

It may be useful to illustrate point 3 above in two trace sector. Let $A$ and $B$ each belong
to single trace cohomology. It is obvious that $AB$ is $Q$ closed. Further, it is always possible to choose our single trace basis such that $A$ and $B$ are $\alpha_1$ and $\alpha_2$ (provided $A$ and $B$ are not proportional). It is then clear that $AB=[Q, O]$ as any nonzero term on the RHS of this equation contains a piece proportional to $[Q, \gamma_i]$ for some $\gamma_i$ and so cannot equal $\alpha_1 \alpha_2$. Consequently $AB$ belongs to $Q$ cohomology. Further if $A$ and $A'$ are $Q$ equivalent it is obvious that $AB$ and $A'B$ are $Q$ equivalent for any $Q$ closed $B$. Finally it is also clear that the operators $\alpha_i \alpha_j$ are $Q$ inequivalent for different values of the pair of
indices $\{i,j\}$. All this establishes that the symmetric product of single trace cohomology lies within the fock space of $Q$ acting on the symmetric product Hilbert space.
Similar arguments may be used to establish the strict equality of these
two constructions.

We now turn the question: What is the cohomology of $Q$ in the
single trace sector? While we are not aware of a complete proof of this
result, it seems overwhelmingly likely that this cohomology is simply
given by the set of $1/16$ BPS descendants of chiral primary operators, i.e.
the list of 1/16 BPS single gravitons in $AdS_5$. Nontrivial evidence for this
conjecture was reported in \cite{Janik:2007pm}. Assuming this to be the case,
we have proved in this section that the cohomology of $Q$ in is given by
the fock space of 1/16 BPS gravitons at energies smaller than $N$. As
we have discussed in the introduction this is already a nontrivial result; it
implies that $1/N$ effects cannot renormalize the energy of a collection of
1/16 BPS gravitons.

At energies larger than $N$ the arguments of this subsection no longer apply;
indeed that is a good thing as the entropy of the fock space of supersymmetric
gravitons grows with energy like  $E^{{\frac{5}{6}}}$, a growth that is
too slow to account for the ${\cal O}(N^2)$ states at energies of order $N^2$
of $1/16$ BPS supersymmetric black holes. In the next section we will begin
a systematic investigation of $Q$ cohomology at energies larger than $N$.
Unfortunately we will be able to report only modest progress in characterizing
this cohomology.

\subsection{Generalities on classical cohomology}
\label{local}
$\mathcal{N}=4$ SYM has 6 scalars $\Phi_{ij}$, 4 chiral
fermions $\Psi_{i\alpha}$ and a gauge field $A_{\alpha\dot{\beta}}$. The scalars $\Phi_{ij}$ are in the
antisymmetric product of $SU(4)$. The scalar fields satisfy
$\Phi_{ij}^*=\Phi^{ij}$ where $\Phi^{ij}={1\over
2}\epsilon^{ijkl}\Phi_{kl}$. The complex conjugates of the fermions
are $\bar{\Psi}^i_{\dot{\alpha}}$.

We will write the field content in terms of chiral fields defined as
(we use the convention $\epsilon^{4mnp}=+\epsilon^{mnp}$):
\begin{equation}\begin{split}
\Phi^{4m}&=\bar{\phi}^m \qquad {1\over
2}\epsilon_{pmn}\Phi^{mn}=\Phi_{4p}=\phi_p \qquad
\Psi_{4\alpha}=\lambda_{\alpha} \qquad
\Psi_{m\alpha}=\psi_{m\alpha}\\
\end{split}\end{equation} with
$m,n,p=1\ldots 3$.

Also, from now on we will denote the special supercharges as
$Q^4_{\alpha}=Q_\alpha$, $S_4^{\alpha}=S^{\alpha}$
and denote the remaining supercharges as $Q^m_{\alpha}$ with $m=1,2,3$.
With these definitions, the action of the
supercharges on the fields is listed in appendix
\ref{transformations}.

To begin with, we consider the cohomology of $Q_-$ at zero
coupling, where all commutators in the supersymmetry algebra vanish. The supersymmetry algebra has:
\begin{equation}\label{bpsrelation}
\Delta\equiv2\{Q_-,S^-\}=E-2J-H_1-H_2-H_3
\end{equation}
where $E$ is the dilatation operator, or the energy in radial quantization,
$J$ is the left $SU(2)$ charge and $H_i$ are the $SU(4)$ Cartans.
At zero coupling, we can solve the cohomology problem by simply listing all basic fields or `letters' in the theory which have $\Delta=0$.
These are $\bar{\phi}^m$, $\bar{\lambda}_{\dot{\alpha}}$, $\psi_{m+}$,
$f_{++}$ and derivatives $D_{+\dot{\alpha}}$ acting on them, where
$f_{++}$ denotes the ${++}$ component of the
field strength $f_{\alpha\beta}$.
A quick look at the supersymmetry algebra shows that these letters are $Q_-$ closed, but not $Q_-$
exact. The gaugino equation of motion,
\begin{equation}
\partial_{+\dot{\beta}}\bar{\lambda}^{\dot{\beta}}=0\ ,
\end{equation}
is the only equation of motion that can be constructed out of these letters.
At zero coupling, any operator constructed out of the $\Delta=0$ letters, modulo this equation of motion,
will be $1/16$ BPS. The partition function over the above letters can easily be calculated and can be
found in \cite{Kinney:2005ej}.

At finite coupling, the commutators appearing on the right hand side of the supersymmetry algebra in appendix \ref{transformations} introduce constraints on the free cohomology.  The essential point in what follows is that we will formulate the supersymmetry algebra in a way that makes some of these constraints easy to implement in some sectors.

The action of the special supercharge $Q_-$ on the supersymmetric
letters is:
\begin{equation}\begin{split}
[Q_-,\bar{\phi}^n]&=0\\
\{Q_-,\psi_{n+}\}&=\epsilon_{nmp}[\bar{\phi}^m,\bar{\phi}^p]\\
\{Q_-,\bar{\lambda}_{\dot{\beta}}\}&=0\\
[Q_-,f_{++}]&=i[\bar{\phi}^m,\psi_{m+}]\ .
\end{split}\end{equation}
We define a field corresponding
to each supersymmetric letter to simplify the analysis of derivatives:
\begin{equation}\label{fields}\begin{split}
\bar{\phi}^m(z)&=\sum_n {z_1^{n_1}z_2^{n_2}\over n_1!n_2!}
D_1^{n_1}D_2^{n_2}\bar{\phi}^m\\
f(z)&=\sum_n{z_1^{n_1}z_2^{n_2}\over n_1!n_2!}
 D_1^{n_1}D_2^{n_2}f_{++} \\
\psi_{m+}(z)&=\sum_n {z_1^{n_1}z_2^{n_2}\over n_1!n_2!}
D_1^{n_1}D_2^{n_2}\psi_{m+}\\
\bar{\lambda}_{\dot{\alpha}}(z)&=\sum_n {z_1^{n_1}z_2^{n_2}\over
n_1!n_2!} D_1^{n_1}D_2^{n_2}\bar{\lambda}_{\dot{\alpha}}\ .
\end{split}\end{equation}
The derivatives $D_{+\dot{\alpha}}$ have been abbreviated as $D_1,D_2$ and all
derivatives should be understood to be symmetrized. That is
$D_1D_2\phi$ denotes ${1\over2}(D_1D_2+D_2D_1)\phi$ and so on. With
these definitions the action of $Q_-$ on the supersymmetric
derivatives is
\begin{equation}
[Q_-,D_{\dot{\alpha}}\zeta]=-i[\bar{\lambda}_{\dot{\alpha}},\zeta]+D_{\dot{\alpha}}
Q_-\zeta\ ,
\end{equation}
and the action of $Q_-$ on the supersymmetric fields is
\begin{equation}\begin{split}\label{Q}
[Q_-,\bar{\phi}^m(z)]&=-i[z^{\dot{\alpha}}(1+z\cdot\partial)^{-1}\bar{\lambda}_{\dot{\alpha}}(z),\bar{\phi}^m(z)]\\
[Q_-,f(z)]&=-i[z^{\dot{\alpha}}(1+z\cdot\partial)^{-1}\bar{\lambda}_{\dot{\alpha}}(z),f(z)]
+i\ [\bar{\phi}^n(z),\psi_{n+}(z)]\\
\{Q_-,\bar{\lambda}_{\dot{\beta}}(z)\}&=-i\{z^{\dot{\alpha}}
(1+z\cdot\partial)^{-1}\bar{\lambda}_{\dot{\alpha}}(z)
,\bar{\lambda}_{\dot{\beta}}(z)\}\\
\{Q_-,\psi_{m+}(z)\}&=-i[z^{\dot{\alpha}}(1+z\cdot\partial)^{-1}\bar{\lambda}_{\dot{\alpha}}(z)
,\psi_{m+}(z)] +\epsilon_{mnp}[\bar{\phi}^n(z),\bar{\phi}^p(z)]\ .
\end{split}\end{equation}
In particular, the action of $Q_-$ has two terms: The first has the form of an infinitesimal
gauge transformation parameterized by the object
$z^{\dot{\alpha}}(1+z\cdot\partial)^{-1}\bar{\lambda}_{\dot{\alpha}}(z)$ and
the second is completely local with these field definitions.

We include here a proof of the above supersymmetry transformations
for fields which are independent of $z_2$. For simplicity of
notation, we write $Q_-= Q$ and $D_1=D$:
\begin{equation}\begin{split}
[Q,\xi(z)]&=z\sum_n\left({z^{n-1}\over n!}
\sum_{k=0}^{n-1}\sum_{l=0}^k {k!\over l!(k-l)!}[D^l\lambda,D^{n-l-1}\xi]\right)+\sum_n{z^n\over n!}D^n[Q,\xi]\\
&=-iz\sum_n\left({z^{n-1}\over n!}
\sum_{l=0}^{n-1}\left(\sum_{k=l}^{n-1} {k!\over l!(k-l)!}\right)
[D^l\lambda,D^{n-l-1}\xi]\right)+\sum_n{z^n\over n!}D^n[Q,\xi]\\
&=-iz\sum_n\left({z^{n-1}\over n!} \sum_{l=0}^{n-1}{n!\over
(l+1)!(n-l-1)!}
[D^l\lambda,D^{n-l-1}\xi]\right)+\sum_n{z^n\over n!}D^n[Q,\xi]\\
&=-iz\sum_n\sum_{l=0}^{n-1}[{D^l\lambda\over (l+1)!}z^l,
{D^{n-l-1}\xi\over (n-l-1)!}z^{n-l-1}]\\
&=-i[z(1+z\partial)^{-1}\lambda(z),\xi(z)]+[Q,\xi](z)
\end{split}\end{equation}
The general case, allowing both derivatives follows by similar arguments which keep track of
the fact that derivatives are symmetrized. We note that it is also possible to write the action of some of the supercharges which commute with $Q_-,S^-$ in a form similar to that in equation \ref{Q}.

\subsection{1/16 BPS Cohomology at Finite $N$ in the Sector Made of Bosonic Operators}
\label{scalarsector}

In this subsection, we will describe the $1/16$ BPS cohomology in the sector
where operators are constructed from any number of derivatives, scalars
and gauge fields. More precisely we will study the counting all elements of
$Q$ cohomology that are $Q$ equivalent to a purely bosonic operator.

Let us first study consequences of the requirement that the operators
we study are $Q$ closed. Recall that the action of $Q$ on $\bar{\phi}^m$
is proportional to the gaugino operators. Since we are considering only operators which themeselves contain no gauginos, we may simply regard
the gaugino field as an arbitrary holomorphic fermionic field. Consequently,
the transformation
\begin{equation}\label{scalar}
Q_-\bar{\phi}^m(z)=-i[z^{\dot{\alpha}}(1+z\cdot\partial)^{-1}\bar{\lambda}_{\dot{\alpha}}(z),\bar{\phi}^m(z)]
\end{equation}
is simply a gauge transformation parameterized by
$W(z_1,z_2)=z^{\dot{\alpha}}(1+z\cdot\partial)^{-1}\bar{\lambda}_{\dot{\alpha}}(z)$.
We conclude that the set of $Q$ invariant operators constructed out of the fields  $\bar{\phi}^m(z)$ is simply the set of operators made out of these
fields that are invariant under $z$ dependent $U(N)$ gauge transformations acting on $\bar{\phi}^m(z)$.

We reiterate that $Q$ closed operators constructed out of $\bar{\phi}^m(z)$ (and,  as we will see below the gauge field $f(z)$) must be gauge invariant under all \textit{holomorphic} gauge transformations $W(z_1,z_2)$. If we choose to construct gauge invariants using traces, the requirement of $Q$ closedness requires that every field inside any given trace is evaluated at the same
$z$. Of course different traces may be evaluated at different values of $z$.
For example, if $x\ne y$, we may consider $\tr \bar{\phi}^1(x)\bar{\phi}^2(x)$,
but not $\tr \bar{\phi}^1(x)\bar{\phi}^2(y)$.

In order to understand the constraints from $Q$ exactness consider
\begin{equation}\label{comm}
\{Q_-,\psi_{m+}(z)\}=-i\{z^{\dot{\alpha}}(1+z\cdot\partial)^{-1}\bar{\lambda}_{\dot{\alpha}}(z)
,\psi_{m+}(z)\} +\epsilon_{mnp}[\bar{\phi}^n(z),\bar{\phi}^p(z)].
\end{equation}
This relation implies that operators containing commutators of scalars inside a trace are $Q_-$ exact. In particular if $|\chi\rangle=\tr
\left(A(z)[\bar{\phi}^m(z),\bar{\phi}^n(z)]\right)\times \ldots$
and $|\chi\rangle'={1\over2}\tr\left(
A(z)\epsilon^{mnk}\psi_{k+}(z)\right)$ then
$Q_-|\chi\rangle'=|\chi\rangle$ (the terms in $Q$ variation that involve
the gaugino cancel out because of the `gauge' invariance of $|\chi\rangle '$
as described in the previous paragraph).

Now let us study the requirement of $Q$ invariance of operators containing the gauge field $f$. We see that the two terms on the right hand side of
\begin{equation}
[Q_-,f(z)]=-i[z^{\dot{\alpha}}(1+z\cdot\partial)^{-1}\bar{\lambda}_{\dot{\alpha}}(z),f(z)]
+i\ [\bar{\phi}^n(z),\psi_{n+}(z)]
\end{equation}
must be annihilated separately since one involves a gaugino and the other a chiralino. The first term is the same
gauge transformation we saw above in equation (\ref{scalar}) and the second constraint ensures that no operator
constructed purely from bosonic letters may contain $f(z)$ except the operator $\tr f(z)$ in the $U(N)$ theory.

Therefore states in $Q$ cohomology in the $SU(N)$ theory that are composed
entirely out of bosonic letters can chosen to
satisfy two constraints: They must be gauge invariant functions
of the local fields $\bar{\phi}^m(z)$ and they can be chosen to be completely
symmetrized on all scalars inside any given trace.

We have so far been studying operators constructed out of the generating
functions $\bar{\phi}^m(z)$. While these generating fields are convenient
for many purposes, they do not carry definite values of the angular momentum
quantum number $J$. If we are interested in counting operators graded by $J$, as we typically are, then we must eventually return to the derivative basis.
 In this basis the single trace operators in $Q$ cohomology are
\begin{equation}\label{ststates}
D_1^{k_1}D_2^{k_2}\tr (\bar{\phi}^1)^{n_1} (\bar{\phi}^2)^{n_2}
(\bar{\phi}^3)^{n_3}\ ,
\end{equation}
where $D_1,D_2$ refer to the derivatives $D_{+\dot{\alpha}}$, which can be
regarded as ordinary derivatives since they act on gauge-invariants. The
scalars inside the trace is regarded as being symmetrized.
The only remaining constraints on the cohomology in this sector are the trace relations which
become important for operators with more than $N$ letters and reduce the number of independent
operators. In the $U(N)$ theory, the field strength $f(z)$
may also participate, but in a rather trivial way as explained above.

This description of the scalar operators in $Q$ cohomology is not yet explicit
enough to provide a simple counting rule to ennumerate these operators. The reason for this is that we have not yet come to grips with the trace identities that complicate this ennumeration. We will have only modest success in
taming these identities in the next section.

\section{Partition Functions at Finite $N$}
\label{partfunctions}

In this section, we will compute the exact partition function of the
$\frac{1}{16}$-BPS cohomology in several subsectors involving
scalars and derivative operators only: the latter restriction meaning that
the cohomology has at least one representative
made of scalars and derivatives only, as discussed in the previous section.

To be concrete, we provide a
partition function of cohomologies involving one species of scalar
and arbitrary derivatives in $U(N)$ gauge theory, and also involving
two species of scalars and arbitrary derivatives in $U(2)$ gauge
theory. The combinatorics of plethystics and syzygies, explored
recently in \cite{Benvenuti:2006qr,Feng:2007ur} in the context of
chiral rings, proves useful in obtaining the partition function of
the two scalar species $U(2)$ subsector.

We also investigate a partition function which provides an upper bound
for the exact degeneracy of $\frac{1}{16}$-BPS cohomology involving all scalars and derivatives.
The latter upper bound is obtained by loosening the condition
for the $\frac{1}{16}$-BPS cohomology and we explain that it has a clear interpretation
in the dual gravity context. The same upper bound partition function can be obtained
by `naively' quantizing the $\frac{1}{16}$-BPS fluctuations on
dual giant gravitons in $AdS_5\times S^5$.
The purpose of studying the upper bound partition function
is twofold. One is to conclusively show that one cannot reproduce the
entropy of supersymmetric black holes from the purely bosonic cohomology.
Another purpose is to try to have a better
understanding of the $1/16$ BPS states in the regime $g_sN\gg 1$,
$E\gtrsim N$ using giant gravitons.

\subsection{Partition Function of a Single Scalar and all Derivatives at Finite $N$}
\label{singlescalar}
Now we will consider the sector generated by a single scalar, say $\bar{\phi}^1$ and all derivatives.
Our prescription states that
the operators\footnote{Such states are in
fact $1/8$ BPS states, but preserve a different set of supercharges from
the better studied chiral ring. The chiral ring (which can be constructed
from the zero modes of the 3 complex scalars, $\bar{\phi}^i$)
counts states that are annihilated by $Q_\alpha$ and $S^\alpha$, but
the states of this subsection are annihilated by $Q_-,S^-$ and $Q^1_-,S_1^-$
instead.} we should count are generated as products of the single trace
states $\tr \left(\bar{\phi}^1(z)\right)^n$.
These operators may be counted very simply at a given $z$ even
accounting for trace identities;
the answer is given simply by all polynomials of
 $\tr \left(\bar{\phi}^1(z)\right)^n$ \cite{Berenstein:2004kk}. In order
to account for the $z$ dependence we must count all polynomials of
\begin{equation}
D_1^{k_1}D_2^{k_2}\tr(\bar{\phi}^1)^n \qquad n=1,\ldots,N
\end{equation}
The multi-trace partition function is given by the formulae of Bose statistics
\begin{equation}\label{gautam}
Z_N(\mu_1,\theta_1,\theta_2)=
\prod_{k_1,k_2=0}^\infty \prod_{n=1}^N{1\over 1-\theta_1^{k_1}\theta_2^{k_2}\mu_1^n}
\end{equation}

We now turn to the bulk interpretation of this partition function. Let us
first study our partition function in terms of giant gravitons.
The giant gravitons which are dual to the operators considered in this
section are $1/8$ BPS D3-branes that wrap $S^3s$ with a particular orientation
on the $S^5$ and move on a given (pointlike) trajectory in $AdS_5$
\cite{Arapoglu:2003ti, Caldarelli:2004yk,Mandal:2006tk}. The set of such
configurations can be quantized and the resulting partition function was
computed in \cite{Mandal:2006tk} and agrees exactly with \ref{gautam}.

It is natural to inquire whether this partition function admits other
complementary bulk interpretations.
Recall that the BPS sector of Yang-Mills chiral ring admited two
complementary quantizations; the first  \cite{Beasley:2002xv,Biswas:2006tj}
from quantizing gravitons \cite{Mikhailov:2000ya}, and the
second quantizing by quantizing dual giant gravitons
\cite{Mandal:2006tk}. (The latter
approach has been generalized \cite{Martelli:2006vh,Basu:2006id} to
certain $\mathcal{N}=1$ superconformal theories.) The two
descriptions yield exactly the same partition function in this
sector, which also agrees with the Yang-Mills theory result
\cite{Kinney:2005ej}. In section \ref{threefour} we will discuss problems with
an analogeous dual giant interpretation of \eqref{gautam}

\subsection{Exact Partition Function in Sector with 2 Scalars for
$U(2)$}\label{utwopartition}

In the last subsection we presented an exact partition function over operators in $Q$ cohomology that are composed of arbitrary numbers of derivatives of a single
scalar field. In this subsection we will present the only other exact partition function in our paper - the partition function over operators in $Q$ cohomology that are composed of arbitrary numbers of derivatives of two scalar fields - but only for the gauge group $U(2)$.

Restated, in this section, we will calculate explicitly the partition
function in the sector with 2 scalars, and all derivatives for
gauge group $U(2)$. For convenience, we denote the 2 scalars as $X,Y$. We will employ the plethystic
exponential described in \cite{Benvenuti:2006qr,Feng:2007ur}.
An independent check of our partition function can be found
in appendix \ref{checks}.

We first consider multi trace states constructed from
the two scalars $X,Y$ without any derivatives.
These are $1/4$ BPS states and their counting is well known; see for example
\cite{Kinney:2005ej}. At finite $N$, one must restrict the total number of traces used to form independent
multi-trace operators to be less or equal to $N$. For convenience, we include a trivial identity operator
${1\over N}{\rm tr}({\bf 1}_N)$ and constrain the number of traces
to be exactly $N$. The partition function over the $1/4$ BPS states
in the $U(N)$ theory is the coefficient of $p^N$ in
\begin{equation}
  Z(p,\mu_1,\mu_2)\equiv\sum_{N=0}^\infty p^N
  Z_N(\mu_1,\mu_2)=\prod_{n_1,n_2=0}^\infty
  \frac{1}{1-p\ \mu_1^{n_1}\mu_2^{n_2}}\ .
\end{equation}
The variable $p$ is a chemical potential for the number of traces.
In particular, the $\frac{1}{4}$-BPS partition function for the
$U(2)$ theory can be can be obtained by computing
$Z_2=\left.\frac{1}{2}\frac{\partial^2}{\partial
p^2}Z(p,\mu_1,\mu_2)\right|_{p=0}$:
\begin{equation}\label{quarter}
  Z_2(\mu_1,\mu_2)=\frac{1+\mu_1\mu_2}{(1-\mu_1)^2(1-\mu_2)^2
  (1+\mu_1)(1+\mu_2)}\ .
\end{equation}

An alternative way of imposing the trace relation constraint is to
leave the number of traces unrestricted, but to instead restrict the
number of letters inside each trace, as discussed in section \ref{singlescalar}.
For gauge group $U(2)$, the following single trace operators generate the most
general multi-trace operators in the $1/4$ BPS sector:
\begin{equation}
  \mathcal{O}_{n_1n_2}\equiv{\rm tr}\left(X^{n_1}Y^{n_2}\right)\ ,\ \ n_1+n_2\leq 2\ ,
\end{equation}
They are called primitive operators, or generators.
While these primitive
operators generate all multi-trace operators in the $1/4$ BPS sector,
there is a redundancy, or overcounting of
operators since there may be more than one polynomial relation
among $\mathcal{O}_{n_1n_2}$ arising from trace relations. Such
relations are called \textit{syzygies} \cite{Feng:2007ur}. Explicitly, in
the $\frac{1}{4}$-BPS sector with $U(2)$ gauge group, there are five
primitive operators
\begin{equation}
  \mathcal{O}_{10}={\rm tr}(X)\ ,\ \ \mathcal{O}_{01}={\rm tr}(Y)\ ,\ \ \mathcal{O}_{20}={\rm tr}(X^2)\ ,\ \
  \mathcal{O}_{11}={\rm tr}(XY)\ ,\ \ \mathcal{O}_{02}={\rm tr}(Y^2)\ ,
\end{equation}
which are subject to only one syzygy (see section 6 of \cite{Feng:2007ur})
\begin{equation}\label{syz}
  \mathcal{O}_{20}(\mathcal{O}_{01})^2+\mathcal{O}_{02}
  (\mathcal{O}_{10})^2+2(\mathcal{O}_{11})^2-2\mathcal{O}_{20}
  \mathcal{O}_{02}-2\mathcal{O}_{10}\mathcal{O}_{01}
  \mathcal{O}_{11}=0\ .
\end{equation}
One may regard this syzygy as relating $(\mathcal{O}_{11})^2=({\rm
tr}(XY))^2$ to a combination of primitive operators containing no
more than one $\mathcal{O}_{11}$.

The partition function of the primitive single trace operators is
\begin{equation}
  z^\prime_2(\mu_1,\mu_2)\equiv\mu_1+\mu_2+\mu_1^2+\mu_1\mu_2+\mu_2^2\ .
\end{equation}
Had one been ignoring the
syzygy, we would have obtained a multi-trace partition function
simply by multiparticling (or taking plethystic exponential of)
this single trace partition function
\begin{equation}
  Z^\prime_2(\mu_1,\mu_2)=\exp\left(\sum_{r=1}^\infty\frac{z^\prime_2
  (\mu_1^r,\mu_2^r)}{r}\right)=\frac{1}{(1-\mu_1)(1-\mu_2)(1-\mu_1^2)
  (1-\mu_2^2)(1-\mu_1\mu_2)}\ .
\end{equation}
Comparing with the correct answer $Z_2(\mu_1,\mu_2)$, we find that
$Z^\prime_2$ overcounts the states, simply because to the relation (\ref{syz})
is ignored. subtracting this, the correct partition function should be
\begin{equation}\label{subtractsyzygy}
  Z_2(\mu_1,\mu_2)=Z^\prime_2(\mu_1,\mu_2)-\mu_1^2\mu_2^2
  Z^\prime_2(\mu_1,\mu_2)=(1-\mu_1^2\mu_2^2)Z^\prime_2(\mu_1,\mu_2)\ ,
\end{equation}
which is indeed true. The second subtracted term corresponds to
eliminating the contribution to the partition function from
operators of the form
\begin{equation}
  \left(\frac{}{}{\rm tr}(XY)\right)^2
  \left(\frac{}{}{\rm arbitrary\ multiplication}\frac{}{}\right)\ ,
\end{equation}
since we do not want to count any operators containing more than one
$\mathcal{O}_{11}={\rm tr}(XY)$.

The above elimination of overcounting, or compensation for syzygies,
can be phrased in terms of the single trace partition function. The factor
$(1-\mu_1^2\mu_2^2)$ in (\ref{subtractsyzygy}) can be regarded as
coming from the Plethystic exponential
of $-\mu_1^2\mu_2^2$. In other words, the `effective' single trace
partition function which gives the correct multi-trace answer is
\begin{equation}\label{singletraceU(2)}
  z_2(\mu_1,\mu_2)=\mu_1+\mu_2+\mu_1^2+\mu_1\mu_2+\mu_2^2-\mu_1^2\mu_2^2\ .
\end{equation}
Formally, the last term may be regarded as eliminating the
redundant generator $({\rm tr}(XY))^2$ by giving it `degeneracy' $-1$.

Now we wish to add derivatives to each single trace operator. First,
to each term in the single-trace partition function with positive
coefficients (corresponding to a primitive operator), we multiply a
factor
\begin{equation}
  \frac{1}{(1-\theta_1)(1-\theta_2)}\equiv
  \sum_{k_1,k_2=0}^\infty(\theta_1)^{k_1}(\theta_2)^{k_2}\ ,
\end{equation}
where $\theta_1$ and $\theta_2$ are chemical potentials for the two
angular momenta, i.e., numbers of derivatives. Furthermore, the form
of states that should be eliminated (due to syzygies) should also be
multiplied by this factor, since any set of derivatives acting on a
syzygy also represents a redundancy to be eliminated from the
counting. Therefore, one obtains the following single-particle and
multi-particle partition functions:
\begin{eqnarray}
  z_2(\mu_1,\mu_2,\theta_1,\theta_2)&=&
  \frac{z_2(\mu_1,\mu_2)}{(1-\theta_1)(1-\theta_2)}\\
  Z_2(\mu_1,\mu_2,\theta_1,\theta_2)&=&\exp\left(\sum_{r=1}^\infty
  \frac{z_2(\mu_1^r,\mu_2^r,\theta_1^r,\theta_2^r)}{r}\right)\ .
\end{eqnarray}
Following the above procedure, one finds
\begin{equation}\label{utwo}
 \hspace{-1.5cm} Z_2(\mu_1,\mu_2,\theta_1,\theta_2)=\prod_{k_1,k_2=0}^\infty
  \frac{1-\mu_1^2\mu_2^2\theta_1^{k_1}\theta_2^{k_2}}{(1-\mu_1\theta_1^{k_1}\theta_2^{k_2})
  (1-\mu_2\theta_1^{k_1}\theta_2^{k_2})(1-\mu_1^2\theta_1^{k_1}\theta_2^{k_2})(1-
  \mu_2^2\theta_1^{k_1}\theta_2^{k_2})(1-\mu_1\mu_2\theta_1^{k_1}\theta_2^{k_2})}\ .
\end{equation}
In the appendix \ref{checks} we provide an independent nontrivial
check of this result.

It is natural to wonder whether the $U(N)$ partition function
including all three scalars and all derivatives may be explicitly
calculated by the same procedure using the single trace basis.
With $U(2)$ group, an explicit check of the kind carried out in
the appendix \ref{checks}, which can be easily extended to the
cases with three scalars,
shows that this is \textit{not} the case. We find that an apparent
reason for this failure seems to be the following: while the single
trace partition function (\ref{singletraceU(2)}) is a finite series
with two scalars, the similar function becomes an infinite series
with three scalars in $U(2)$. This is pointed out to correspond to the fact
that $\mathbb{C}^4/\mathbb{Z}_2$ is a `complete intersection,' while
$\mathbb{C}^6/\mathbb{Z}_2$ is not \cite{Feng:2007ur}. There could
be more delicate combinatoric structures applicable to all cases,
beyond what we found in the case with $U(2)$ two scalars.

\subsection{Upper bound on scalar sector and high energy scaling}
\label{upperbound}

While we have not been able to find an exact formula to count all operators
in Q cohomology that admit representatives composed purely of scalar fields,
in this subsection we will present a rigorous upper bound for the growth
in the number of such operators as a function of their energy.
In particular we will demonstrate that when the charges and energy of operators
are taken to be $O(N^2)$, this sector does not contain enough operators to reproduce the $N^2$
scaling of the entropy of black holes in $AdS_5\times S^5$.

Firstly, for the purpose of enumeration, the scalars in our cohomology
may be regarded as commutating, or
diagonal, matrices. Our upper bound is obtained in the $U(N)$
theory by allowing any number of
derivatives to act on any of the $N$ eigenvalues of any of the three scalars.
That is, we will ignore the fact that
the eigenvalues should really be symmetrized \textit{inside} each trace, before
the action of the derivatives.
Counting this larger set of operators will give us
an upper bound for the number of operators in the scalar and derivatives sector.

Let us first characterize the above `relaxed' Hilbert space.
This simply consists of the $N$ bosonic particle states made of
the single eigenvalue Hilbert space, which we call $\mathcal{H}_1$.
The last $\mathcal{H}_1$
is characterized as follows:  We have the letters
\begin{equation}
\partial_1^{k_1}\partial_2^{k_2}x_i \qquad k_j=0,\ldots,\infty,\ i=1,2,3
\end{equation}
and $\mathcal{H}_1$ is made of all words constructed from these
commuting letters.  Then the full partition function is the
coefficient of $p^N$ in
\begin{equation}
Z=\prod_{n^i_k=0}^\infty{1\over 1-p e^{-\beta\sum_{k,i}n^i_k
(1+k^i_1+k^i_2)}}\ ,
\end{equation}
where the $k^i_j=0,\ldots,\infty$ and $i=1,2,3$ and $j=1,2$. (We have set
$e^{-\beta}=\mu_i=\theta_a$ for simplicity.)

Let us pause for a brief comment. As we have described above, the limit $N \to \infty$ our
relaxed Hilbert space is given by the Fock space the Hilbert space of supersymmetric scalar states of $U(1)$ Yang Mills theory. On the other hand
the restriction to scalars of the correct $N\to \infty$ limit of the 1/16 BPS partition function (see section 2) is given by the Fock space of a much
smaller Hilbert space - the space of scalar 1/16 BPS descendents of chiral
primaries. This makes clear that our relaxed Hilbert space is much larger
than the actual Hilbert space a energies small compared to $N$. On the other
hand we argue in the next section that the relaxed Hilbert space is not very
different from the exact space at the opposite high energy end.

Now one obtains the free energy and $N$ in terms of $p$ and the
temperature $\beta$:
\begin{equation}\label{N}\begin{split}
F&=-1/\beta \ln Z = {1\over\beta}\sum_{n^i_k=0}^\infty
\ln\left(1-pe^{-\beta\sum_{k,i}n^i_k(1+k_1^i+ k^i_2)}\right)\\
\langle N\rangle &= p{\partial\over\partial p}\ln Z =
\sum_{n^i_k=0}^\infty{1\over p^{-1}e^{\beta\sum n^i_k(1+k_1^i+ k^i_2)}-1}\ .
\end{split}\end{equation}
$p$ must take a value in $[0,1]$. We now consider the high temperature limit
$\beta\ll 1$. In this limit, there are a large number of states such
that $\beta\sum n_k^i(1+ k^i_1+k^i_2) \sim 0$ so that if $p\sim 1$, these
states will produce a divergent contribution to the right hand side
of (\ref{N}). Since the left hand side is finite, it must be that
$p\ll 1$ when $\beta\ll 1$.

In this case, we may approximate the particle number as
\begin{equation}\label{Napprox}\begin{split}
\langle N\rangle\approx
\sum_{n^i_k=0}^\infty pe^{-\beta\sum n_k^i (1+k^i_1+k^i_2)}=p
\prod_{k^i_j} \sum_{n^i_k=0}^\infty e^{-\beta n_k^i(1+k^i_1+k^i_2)} =p
\left[\prod_{k_1,k_2=0}^\infty {1\over 1-e^{-\beta (1+k_1+k_2)}}\right]^3\ .
\end{split}\end{equation}
Next we calculate $\langle E\rangle$:
\begin{equation}\begin{split}
\langle E\rangle &= \sum_{n^i_k=0}^\infty {\sum n^i_k (1+k^i_1+k^i_2) \over
p^{-1}e^{\beta\sum n_k^i (1+k^i_1+k^i_2)}-1}\approx p\sum_{n^i_k=0}^\infty
e^{-\beta\sum n_k^i (1+k^i_1+k^i_2)}\left[\sum n^i_k (1+k^i_1+k^i_2)\right]\\
&=\left[\prod_{k_1,k_2=0}^\infty {1\over 1-e^{-\beta
(1+k_1+k_2)}}\right]^3\left\{3p\sum_{k}{ 1+k_1+k_2 \over e^{\beta
(1+k_1+k_2)}-1}\right\}\\
&=3N\left(\sum_k{1+k_1+k_2 \over e^{\beta (1+k_1+k_2)}-1}\right)\ .
\end{split}\end{equation}

We will next extract the $\beta\rightarrow 0$ asymptotic form of the series
\begin{equation}
  \sum_{k_1,k_2=0}^\infty\frac{1+k_1+k_2}{e^{\beta(1+k_1+k_2)}-1}\ .
\end{equation}
Defining $x\equiv\beta k_1$ and
$y\equiv\beta k_2$, we find that $dx$ and $dy$ are small in the $\beta\rightarrow 0$ limit,
so that we can approximate the series by a 2-dimensional integral. Ignoring $1$ in
the $1+k_1+k_2$ which only gives subleading terms in $\beta$, one obtains
\begin{equation}
  \sum_{k_1,k_2=0}^\infty\frac{1+k_1+k_2}{e^{\beta(1+k_1+k_2)}-1}\approx
  \frac{1}{\beta^3}\int_0^\infty dx\int_0^\infty dy
  \ \frac{x+y}{e^{x+y}-1}\ .
\end{equation}
Defining $t\equiv x+y$ and $u\equiv \frac{x-y}{2}$, this term becomes
\begin{equation}
  \frac{1}{\beta^3}\int_0^\infty dt\int_{-\frac{t}{2}}^{\frac{t}{2}}du\
  \frac{t}{e^t-1}=\frac{1}{\beta^3}\int_0^\infty dt\frac{t^2}{e^t-1}=
  \frac{2\zeta(3)}{\beta^3}\ ,
\end{equation}
where $\zeta(s)\equiv\sum_{k=1}^\infty\frac{1}{k^s}$ is the Riemann's zeta
function. Note that $\zeta(3)=1.2020569\ldots$

The above evaluation gives
\begin{equation}
\langle E\rangle={6N\zeta(3)\over \beta^3}\ .
\end{equation}

Next we process $F$, using the fact that $pe^{-\beta\sum_{k,i}n^i_k
(1+k^i_1+k^i_2)}<<1$, which allows us to truncate the log series:
\begin{equation}
F\approx -{p\over\beta}\sum_{n^i_k=0}^\infty e^{-\beta\sum_{k,i}n^i_k
(1+k^i_1+k^i_2)}=-{p\over\beta}\left[\prod_{k_1,k_2=0}^\infty {1\over 1-e^{-\beta
(1+k_1+k_2)}}\right]^3=-{N\over \beta}\ .
\end{equation}
Since $S=\beta(E-F)\simeq {6\zeta(3)N\over \beta^2}+N$, we have in the limit of small
$\beta$,
\begin{equation}
S={6\zeta(3)N\over \beta^2}.
\end{equation}
Eliminating $\beta$ from the expressions for $\langle E\rangle$ and
$S$, we find that
\begin{equation} \label{grow}
S=\bigg({6\zeta(3)N}\bigg)^{1/3}E^{2/3}.
\end{equation}
So for states with energy $\mathcal{O}(N^2)$, we have $S\sim
N^{5/3}<N^2$. This establishes conclusively that there are not enough
states in this sector to form a black hole.

Note the entropy of a fock space of scalar 1/16 descendants of chiral primaries grows like $K E^{{5/6}}$ where $K$ is a number of order unity. Thus \eqref{grow} has more states than multigravitons for $E\ll N^2$ (this is because relaxation of the Hilbert space greatly increases the number of
states at low energies) but fewer states than multigravitons at $E \gg N^2$
(because \eqref{grow} accounts for finite $N$ truncations absent in the Fock space, which are important at high energies). The two ennumerations yield approximately the same number of states at energies of order $N^2$.

\subsection{Upper bound, exact degeneracy and (dual) giant gravitons}
\label{threefour}

In the $\frac{1}{8}$-BPS sector, the degeneracy of chiral primary
operators has been convincingly reproduced from the bulk perspective
by quantizing giant gravitons
\cite{Mikhailov:2000ya,Beasley:2002xv,Biswas:2006tj,Mandal:2006tk}.
Two complementary approaches are available; one which quantizes the
giant gravitons extended in $S^5$, and another which quantizes the
dual giant gravitons extended in $AdS_5$. In the $\frac{1}{16}$-BPS
sector, or more generally in a sector with nonzero angular momentum
in $AdS_5$, such an understanding from the bulk viewpoint is almost
lacking, except in the very simple sector investigated from the bulk
viewpoint in \cite{Mandal:2006tk} and from gauge theory in our
section 3.1.

The upper bound we provided in the previous subsection has an
interpretation from the bulk perspective. We will show that it can
be regarded as the result of a naive quantization of the
$\frac{1}{16}$-BPS fluctuations \cite{Kim:2006he} on the dual giant
gravitons of \cite{Mandal:2006tk}. As we have already seen in the previous
subsection, the upper bound relaxation is far from a good approximation
of the correct 1/16 BPS Hilbert space at low energies. In this section
we will explain why (from a dual bulk perspective) the naive dual graviton
quantization fails, and also explain when we expect it to yield approximately
reliable results.

First, we briefly review the $\frac{1}{16}$-BPS fluctuations on dual
giant gravitons and interpret the upper bound partition function in
terms of such fluctuations. The giant graviton solutions carrying
angular momenta in $AdS_5$ are constructed in \cite{Kim:2006he},
which takes advantage of the embedding of $AdS_5\times S^5$ in
$\mathbb{C}^{2+1}\times\mathbb{C}^3$ equipped with a flat metric
with two negative signatures. Using the six complex coordinates of
the latter space, the worldvolume of the giant graviton is given by
the 6-dimensional holomorphic subspace defined by 3 complex
equations which are homogeneous in a suitable sense. See
\cite{Kim:2006he} for the details. The latter space becomes
$4$-dimensional as one intersects this 6-dimensional space with
$AdS_5$ and $S^5$. In particular, taking the 3 coordinates of
$\mathbb{C}^3$ and $\mathbb{C}^{2+1}$ to be $x,y,z$ and
$w_1,w_2,w_3$ satisfying $|x|^2+|y|^2+|z|^2=1$ and
$|w_3|^2-|w_1|^2-|w_2|^2=1$, the 3 complex equations for a spherical
$\frac{1}{8}$-BPS dual giant gravitons can be written as
\begin{equation}
  xw_3=c_1\ ,\ \ yw_3=c_2\ ,\ \ zw_3=c_3\ ,
\end{equation}
where $c_1,c_2,c_3$ are constants. Note that one finds that
$|w_3|^2=c_1^2+c_2^2+c_3^2$ is fixed to be a constant, defining a
3-sphere in $AdS_5$. Fluctuations of the above configuration with
angular momenta in $AdS_5$ are given by
\begin{equation}\label{dualfluct}
  w_3x=c_1+f_1(w_1w_3^{-1},w_2w_3^{-1})\ ,\ \
  w_3y=c_2+f_2(w_1w_3^{-1},w_2w_3^{-1})\ ,\ \
  w_3z=c_3+f_3(w_1w_3^{-1},w_2w_3^{-1})
\end{equation}
where $f_i$ are regarded as functions given by Taylor series of the
arguments. As far as one considers small fluctuations from the
spherical dual giants, $|f_i|^2\ll |c_i|^2$, one finds that
$|\omega_3|^2$ is approximately constant and (\ref{dualfluct}) is
nothing but
\begin{equation}
  (x,y,z)\approx c_ie^{it}+
  e^{it}f_i(\hat\omega_1e^{-it},\hat\omega_2e^{-it})
\end{equation}
where $t$ is the time coordinate of the global $AdS_5$ defined by
$w_3=|w_3|e^{-it}$, and the hatted coordiantes satisfying
$|\hat\omega_1|^2+|\hat\omega_2|^2=1$ parametrize $S^3$ worldvolume
of a dual giant.

Ignoring any possible correction as these fluctuations becomes large,
one finds the following partition function for $N^\prime$ dual
giant gravitons with fluctuations: firstly, the 1-particle (or a
single dual giant) partition function is given by
\begin{eqnarray}\label{singledual}
  \widetilde{Z}_1(\mu_i,\theta_a)&=&
  \prod_{j_1,j_2=0}^\infty\frac{1}{(1-\mu_1\theta_1^{j_1}
  \theta_2^{j_2})(1-\mu_2\theta_1^{j_1}\theta_2^{j_2})(1-\mu_3\theta_1^{j_1}
  \theta_2^{j_2})}\\
  &=&\sum_{\{n^i_{j_1,j_2}\}=0}^\infty \prod_{i=1}^3\left(
  \mu_i^{\sum_{j_1,j_2=0}^\infty n^i_{j_1,j_2}}
  \theta_1^{\sum_{j_1,j_2=0}^\infty j_1 n^i_{j_1,j_2}}
  \theta_2^{\sum_{j_1,j_2=0}^\infty j_2 n^i_{j_1,j_2}}\right)\ .\nonumber
\end{eqnarray}
The partition function of $N^\prime$ or less identical dual giant
gravitons is simply its multi-particling:
\begin{eqnarray}\label{longupper}
  \widetilde{Z}(p,\mu_i,\theta_a)&=&\sum_{N^\prime=0}^\infty p^{N^\prime}
  \widetilde{Z}_{N^\prime}(\mu_i,\theta_a)\\
  &=&\prod_{\{n^i_{j_1,j_2}\}=0}^\infty\left(
  1-p\prod_{i=1}^3\left(
  \mu_i^{\sum_{j_1,j_2=0}^\infty n^i_{j_1,j_2}}
  \theta_1^{\sum_{j_1,j_2=0}^\infty j_1 n^i_{j_1,j_2}}
  \theta_2^{\sum_{j_1,j_2=0}^\infty j_2 n^i_{j_1,j_2}}\right)\right)^{-1}
  \ ,\nonumber
\end{eqnarray}
or
\begin{equation}
  \widetilde{Z}_{N^\prime}(\mu_i,\theta_a)=
  \sum_{\{p_k\},\ \sum kp_k=N^\prime}\prod_{k=1}^\infty
  \frac{\widetilde{Z}_1(\mu_i^{\ k},\theta_a^{\ k})^{p_k}}{p_k!k^{p_k}}\ .
\end{equation}
The latter expression can be obtained by explicitly expanding the
so-called Plethystic exponential containing the parameter $p\hspace{2pt}$: see,
for instance, eqn. (2.9) in \cite{Feng:2007ur}. The expression (\ref{longupper})
is what we used to evaluate the upper bound in the
previous section (setting $e^{-\beta}=\mu_i=\theta_a$).

Now we consider the simple subsector studied in section
\ref{singlescalar}. One can think of the states in this sector as
$\frac{1}{8}$ BPS excitations of $\frac{1}{2}$ BPS states which
carry two angular momenta $J_1$, $J_2$ in $AdS_5$, in addition to
the charge $H_3$ conjugate to $\mu_3\equiv \mu$. As mentioned in
section 3.1, the same partition function has been obtained by
quantizing the $\frac{1}{2}$ BPS giant gravitons with
excitations\footnote{Giant gravitons with such excitations are in
fact ${1\over8}$ BPS as discussed in section \ref{singlescalar}.}
carrying nonzero angular momenta $J_1$, $J_2$ \cite{Mandal:2006tk}.
We now consider the subtleties involved in quantizing the
$\frac{1}{2}$ BPS `dual' giant gravitons with excitations carrying
nonzero $J_1$, $J_2$.

The probe dual giant graviton description is reliable only for large
$N$, or more precisely for $N^\prime \ll N$ where $N^\prime$ is the
number of dual giant gravitons. Following
\cite{Suryanarayana:2004ig}, we explore the dual giant graviton
interpretation of BPS states for the cases in which this condition
is not obeyed, and consider the case with $N^\prime \sim N$ or with
small values of $N=1,2,3\ldots$, assuming that supersymmetry will
help. Indeed, going beyond these limits, a prescription is found
that the BPS states in $U(N)$ gauge theory should come from $N$ or
less multiple dual giant gravitons
\cite{Suryanarayana:2004ig,Mandal:2006tk}. This is a prescription
that we shall also assume in the sector we study. With this
assumption, $N^\prime$ in the naive partition function of the
previous paragraph is identified as $N$ for the $U(N)$ gauge theory.
In the simplest case, $N=1$, the naive partition function of a
single dual giant graviton (\ref{singledual}), with nonzero angular
momenta, is simply
\begin{equation}
  \widetilde{Z}_1(\mu,\theta_1,\theta_2)=\prod_{j_1,j_2=0}^\infty\frac{1}
  {1-\mu\theta_1^{j_1}\theta_2^{j_2}}\ ,
\end{equation}
which is exactly the same as that of a $U(1)$ (non-interacting)
Yang-Mills theory.

For $N=2$, we denote the exact, gauge theory partition function as $Z_2(\mu,\theta_a)$
and find that $\widetilde{Z}_2(\mu,\theta_a)$ and $Z_2(\mu,\theta_a)$
start to disagree. For simplicity, let us consider the states with
$J_2=0$, or equivalently, states in the gauge theory with one kind
of derivative only. A series expansion shows that
\begin{eqnarray}
  Z_2(\mu,\theta)&=&\prod_{j=0}^\infty\frac{1}{(1-\mu\theta^j)(1-\mu^2\theta^j)}\\
  \widetilde{Z}_2-Z_2&=&\prod_{j=0}^\infty\frac{1}{1-\mu\theta^j}
  \left(\frac{}{}\theta^2\mu^2+\theta^3(\mu^2+\mu^4)+
  \theta^4(2\mu^2+2\mu^4+\mu^6)+\cdots\right)\
  .\label{su2difference}
\end{eqnarray}
We have left the overall $Z_1(\mu,\theta)$ factor unexpanded, which
is the contribution from the decoupled overall $U(1)$ mode.
Therefore the series inside the parenthesis of (\ref{su2difference})
may be regarded as a result in the $SU(2)$ gauge theory. A curious
point we would like to emphasize is that, the coefficient of
$\theta^j$ in the series is a finite polynomial which does not
receive contributions from $\mu^{2j}$ or higher order terms.
Therefore, considering the partition function of the $SU(2)$ theory
(which is also meaningful in the $U(2)$ theory since the overall
$U(1)$ degrees never interact with others), the upper bound
$\tilde{d}(H_3,J_1)$ (coefficient of $t^{H_3}\theta^{J_1}$) is equal
to the true degeneracy $d(H_3,J_1)$ if $H_3\geq 2J_1$ is satisfied.

We would like to interpret the above (dis)agreement from the
viewpoint of dual giant gravitons. Firstly, each of the two
$\frac{1}{2}$ BPS dual giant gravitons can be described by a complex
variable, say $z$ introduced above, moving in a 2-dimensional
harmonic potential \cite{Mandal:2006tk}. The BPS fluctuations on
these dual giant gravitons with nonzero angular momentum $J_1$
explained above can be regarded as a spherical harmonics expansion
of $z=\sum_{j=0}^\infty z_j Y_{j0}$ in the basis
$Y_{j0}\sim(\hat\omega_1)^j$. Inserting this expansion in the
world-volume action, the harmonic potential for the modes $z_j$
becomes
\begin{equation}
  \frac{1}{2}\sum_{j=0}^\infty\left(|z_j|^2+j(j+2)|z_j|^2\right)=
  \frac{1}{2}\sum_{j=0}^\infty(j+1)^2|z_j|^2\ .
\end{equation}
For two dual giant gravitons and their fluctuations, one has two
towers of modes $z_j^1$ and $z_j^2$. One can separate the dynamics
of the modes $\frac{z_j^1+z_j^2}{2}$, which gives the overall
$U(1)$ partition function factor $Z_1(\mu,\theta)$ in
$\widetilde{Z}_2$, and the modes of relative separation $z_j\equiv
z_j^1-z_j^2$. We concentrate on the latter part. After naive quantization,
the charges coming from the relative motion degrees of freedom are,
\begin{equation}
  H_3=\sum_j N_j\ ,\ \ J_1=\sum_j j N_j\ \ \ \
  (N_j\equiv\frac{j+1}{2}|z_j|^2)\ .
\end{equation}
From the above expressions, we now argue that $H_3\sim J_1$ is the
region in charge space where the two world-volumes of the dual giant
gravitons become likely to intersect due to the fluctuations. To see
this, we note that, with fixed angular momentum $J_1$, the
fluctuation of the relative separation (governed by $z_j$ with
nonzero $j$) becomes biggest if we assign more occupations to the
modes with lowest nonzero value of $j$, namely $j=1$. The case with
biggest fluctuation is thus obtained by assigning $N_1=J_1$,
$N_0=H_3-N_1=H_3-J_1$ and all other $N_j$'s zero. The separation
$|z_0|$ of two dual giants averaged over $S^3$, and its fluctuation
$|z_1|$ are given by
\begin{equation}
  |z_0|=\sqrt{2N_0}=\sqrt{2(H_3-J_1)}\ ,\ \
  |z_{1}|=\sqrt{N_1}=\sqrt{J_1}\ .
\end{equation}
Demanding $|z_0|\gtrsim |z_1|$ for the two worldvolumes not to
intersect, one finds the condition $H_3\gtrsim J_1$ which is
qualitatively similar to the condition $H_3\geq 2J_1$ for the upper
bound to be exact. Therefore, we interpret that our naive
quantization of multiple dual giant gravitons becomes invalid as a
dual giant graviton becomes close to, or intersects (in a
singular way) with another.

The $SU(2)$ example above is rather special since there are two
charges $H_3$ and $J_1$ which can be used to control both the
average separation $r_0$ of two branes and the fluctuation of its
relative separation. For $N\geq 3$, it becomes impossible to control
all the relative separations with these two conserved charges only.
Indeed, as we will see shortly, we could not identify any region in
the charge space where the upper bound becomes exact (except
for some exceptional and rather occasional values of charges).
However, we would still like to argue that our upper bound becomes
approximately valid in certain regimes of charges in which the dual
giant gravitons are less likely to intersect.

For the cases with $N\geq 3$, we still consider the regime $H_3\gg
1$. In the exact partition function $Z(\mu,\theta)$ and the upper bound
partition function
$\widetilde{Z}(\mu,\theta)$, the asymptotic growth of degeneracy as
a function of the charge $H_3$ is captured by studying the degree of
the pole of the partition function as $\mu\rightarrow 1$. Namely,
the Taylor series expansion
\begin{equation}
  \frac{1}{(1-\mu)^\alpha}=\frac{1}{(\alpha\!-\!1)!}
  \left(\frac{d}{d\mu}\right)^{\alpha-1}
  \sum_{n=0}^\infty \mu^n=\frac{1}{(\alpha\!-\!1)!}\sum_{n=0}^\infty
  \frac{(n\!+\!\alpha\!-\!1)!}{n!}\mu^n
\end{equation}
grows like $\sim n^{\alpha-1}\mu^n$ for terms with large $n$, while
the other $\mu$-dependent functions without poles, in our case,
turn out to contribute to the coefficient with alternating signs.
Therefore, with poles of degree $\alpha$ in the partition function,
the asymptotic growth of the degeneracy with large charge $H_3$
should grow like $(H_3)^{\alpha-1}$. \footnote{The same conclusion
can be obtained by the saddle point method, showing that the saddle
point is at $H_3\approx\frac{\alpha}{1-\mu}$ when $H_3\gg 1$.}

From a simple computation, one can show that
\begin{eqnarray}
  \hspace*{-0.5cm}Z_N(\mu,\theta)&=&\frac{Z_1(\mu,\theta)}{(1-\mu)^{N-1}}
  \left(\frac{1}{N!}\prod_{j=1}^\infty\frac{1}{(1-\theta^j)^{N-1}}\right)\\
  &&\hspace{2cm}\times\left(1+(1-\mu)
  \left(\frac{N(N-1)}{4}-\frac{(N+2)(N-1)}{2}\sum_{j=1}^\infty
  \frac{\theta^j}{1-\theta^j}\right)+O(1\!-\!\mu)^2\right)\nonumber\\
  \hspace*{-0.5cm}\widetilde{Z}_N(\mu,\theta)&=&\frac{Z_1(\mu,\theta)}{(1-\mu)^{N-1}}
  \left(\frac{1}{N!}\prod_{j=1}^\infty\frac{1}{(1-\theta^j)^{N-1}}\right)\\
  &&\hspace{2cm}\times\left(1+(1-\mu)
  \left(\frac{N(N-1)}{4}\prod_{j=1}^\infty\frac{1-\theta^j}{1+\theta^j}
  -(N-1)\sum_{j=1}^\infty
  \frac{\theta^j}{1-\theta^j}\right)+O(1\!-\!\mu)^2\right)\nonumber
\end{eqnarray}
where $Z_1(\mu,\theta)=\prod_{j=0}^\infty\frac{1}{1-\mu\theta^j}$ is
again the contribution from the overall $U(1)$ part which we shall
ignore in the arguments below. From these expansions we see that,
for large $H_3$ (or equivalently near $\mu\approx 1^-$), the two
degeneracies (or the partition functions) are asymptotically the
same. Both degeneracies grow as $\sim (H_3)^{N-2}f(J)$, while their
difference grows more slowly as $\sim (H_3)^{N-3}g(J)$ since
\begin{equation}
  \widetilde{Z}_N-Z_N=
  \frac{Z_1(\mu,\theta)}{2(N\!-\!2)!(1-\mu)^{N-2}}\prod_{j=1}^\infty
  \frac{1}{(1-\theta^j)^{N-1}}
  \left(\sum_{j=1}^\infty\frac{\theta^j}{1-\theta^j}-\frac{1}{2}+\frac{1}{2}
  \prod_{j=1}^\infty\frac{1-\theta^j}{1+\theta^j}\right)+\cdots\ (>0)\ .
\end{equation}
The $J$-dependent functions $f(J)$ and $g(J)$
are determined from the $\theta$ dependent coefficients in the above expansions.

We argue that the above scaling behaviors can be qualitatively
reproduced from the dynamics of multiple dual giant gravitons. Again
we keep $J_1$ to be much smaller than $H_3\gg 1$. The $H_3$
dependent part of the degeneracy $d(H_3,J_1)$ will then be simply
determined by the dynamics of $\frac{1}{2}$-BPS dual giant
gravitons, while the $J$-dependent factor can be determined by
investigating the fluctuations on the $\frac{1}{2}$-BPS dual giants.
The phase space of $N$ $\frac{1}{2}$-BPS giant gravitons is an
$N$-th symmetric product of $\mathbb{C}$. Semiclassically, the
number of states with fixed charge
$2H_3=|\vec{r}_1|^2+|\vec{r}_2|^2+\cdots+|\vec{r}_N|^2$ is obtained
by computing the volume of the region in the phase space with fixed
$H_3$. Taking into account the $SU(N)$ condition, which keeps the
contribution from the relative seperation $\vec{r}_i-\vec{r}_j$
only, the volume is proportional to
\begin{equation}
  d(H_3,J)\sim\int d^2\vec{r}_1\cdots d^2\vec{r}_{N-1}
  \delta\left(2H_3-\sum_{i=1}^N|\vec{r}_i|^2\right)\sim (H_3)^{N-2}\ .
\end{equation}
On the other hand, the naive dual giant graviton counting should go
significantly wrong as the two dual giant gravitons become close,
namely, $|\vec{r}_i-\vec{r}_j|^2\lesssim\epsilon(J)$ for any pair
$i,j=1,\cdots,N$ and certain $\epsilon(J)$ whose value is roughly
given by the fluctuations of $S^3$. The volume of the region with
close enough dual giant gravitons should go as
\begin{equation}
  \tilde{d}(H_3,J_1)-d(H_3,J_1)\sim\int d^2\vec{r}_1\cdots d^2\vec{r}_{N-1}
  \delta\left(2H_3-\sum_{i=1}^N|\vec{r}_i|^2\right)
  \theta\left(\epsilon(J_1)-|\vec{r}_i-\vec{r}_j|^2\right)\sim
  \epsilon(J_1)(H_3)^{N-3}\ .
\end{equation}
where $\theta(x)$ is the step function $\theta(x)=1$ for $x>0$ and
zero otherwise. Thus, the qualitative feature of the error in the
naive dual giant counting, described in the previous paragraph, is
the same as what we expect from the intersection of dual giants.

From the above examples -- namely the exactness of $U(1)$ result,
exactness of $SU(2)$ result in certain regime, and also the
semiclassical error estimate for $N\geq 3$ -- it seems that the
overcounting one gets by naively quantizing dual giant gravitons has
to do with ignoring their intersections as fluctuations become
large. Note also that, in the giant graviton counting of
\cite{Mandal:2006tk} which gave us the correct answer, a giant
graviton never intersects with another. It would be interesting to
see if the dual giant graviton like counting is available which
should modify our naive upper bound partition function. For
instance, it may be possible that restricting the world-volume of
multiple dual giants to intersect smoothly, in a suitable sense,
could give the correct answer $Z_N(\mu,\theta)$. Otherwise, it could
turn out that non-Abelian physics would be important as the dual
giant gravitons intersect.

Before concluding this subsection, let us mention a few results for
the $SU(2)$ case. In this case, even for the operators including all
three scalasrs, our upper bound is saturated for large enough
internal charges $H_1$, $H_2$, $H_3$. Namely, we find that
$\max(H_1,H_2,H_3)\geq 2J_1$ is a sufficient condition for the upper
bound $\tilde{d}(H_i,J_1)$ to be exact. This claim can be shown
analytically by induction: assuming that the claim is proved for
operators in which no more than $k$ derivatives act on same
eigenvalues, it is easy to check that the same claim can be shown
for operators in which no more than $k+1$ derivativese act on same
eigenvalues. The claim for $k=1$ is easily proved from the condition
$\max(H_1,H_2,H_3)\geq 2J_1$: for instance if $H_1$ is the largest,
one can rewrite $\partial x$, $\partial y$, $\partial z$ into
$\partial(x^2)$, $\partial(xy)$, $\partial(xz)$, respectively, using
abundant $x$'s in the operator. By investigating the case in which
an exact partition function is available, we find that the condition
$\max(H_1,H_2,H_3)\geq 2J_1$ is indeed a sufficient but may not be a
necessary condition for the upper bound to be exact. For example, in
the $SU(2)$ two-scalar case in which we have an exact partition
function, one finds:
\begin{eqnarray}
   \widetilde{Z}_2(\mu_i,\theta)-Z_2(\mu_i,\theta)&=&\theta(\mu_1\mu_2)+\theta^2
   (\mu_1^3\mu_2+\mu_1^2\mu_2^2+\mu_1\mu_2^3+\mu_1^2+2\mu_1\mu_2+\mu_2^2)\\
   &&+\theta^3(\mu_1^5\mu_2+\mu_1^4\mu_2^2+\mu_1^3\mu_2^3+\mu_1^2\mu_2^4+\mu_1\mu_2^5\nonumber\\
   &&\hspace{0.9cm}+\mu_1^4+3\mu_1^3\mu_2+3\mu_1^2\mu_2^2+3\mu_1\mu_2^3+\mu_2^4+
   \mu_1^2+3\mu_1\mu_2+\mu_2^2)\nonumber\\
   &&+\theta^4(\mu_1^7\mu_2+\mu_1^6\mu_2^2+\mu_1^5\mu_2^3+\mu_1^4\mu_2^4+\mu_1^3\mu_2^5+
   \mu_1^2\mu_2^6+
   \mu_1\mu_2^7+\nonumber\\
   &&\hspace{0.9cm}+\mu_1^6+3\mu_1^5\mu_2+4\mu_1^4\mu_2^2+4\mu_1^3\mu_2^3+4\mu_1^2\mu_2^4+
   3\mu_1\mu_2^5+\mu_2^6\nonumber\\
   &&\hspace{0.9cm}+2\mu_1^4+6\mu_1^3\mu_2+7\mu_1^2\mu_2^2+6\mu_1\mu_2^3+2\mu_2^4+2\mu_1^2
   +4\mu_1\mu_2+2\mu_2^2)+O(\theta^5)\ .\nonumber
\end{eqnarray}
For instance, terms like $\theta^3\mu_1^4\mu_2^4$ or
$\theta^4\mu_1^7\mu_2^3$ which do not satisfy the bound
$\max(H_1,H_2)\geq 2J_1$ and are not forbidden by gauge-invariance,
happen to be zero. It would be interesting to find a stronger bound
which admits a better geometric interpretation in terms of dual
giant gravitons.

\section{$\frac{1}{16}$-BPS Classical Configurations of $\mathcal{N}=4$ Yang-Mills}
\label{five}

In the rest of this paper we switch gears rather abruptly. Instead of pursuing more sophisticated countings of $Q$ cohomology in larger subsectors, we attempt to find a physical interpretation of the mathematically motivated
generating functions, $\bar{\phi}^m(z)$ that we introduced in section \ref{cohomologyapproach} and have used intensively since then. For this purpose we take a step backwards; we turn off all fermions and study classical ${\cal N}=4$ Yang Mills theory.
We will derive a set of
Bogomolnyi type equations for the bosonic $\frac{1}{16}$-BPS
configurations in $\mathcal{N}=4$ Yang-Mills theory. We will analyze
these equations in perturbation theory, propose a quantization of
the bosonic $\frac{1}{16}$-BPS configurations and find that we
recover a characterization equivalent to that of the $1/16$ BPS
cohomology found in section \ref{scalarsector}. In this section, we
will derive the BPS equations in a convenient gauge. Similar
analysis and quantization have been studied in a simpler
$\frac{1}{4}$- and $\frac{1}{8}$-BPS sector of M5 and M2 brane
CFT's, respectively \cite{Bhattacharyya:2007sa}.

\subsection{$\frac{1}{16}$-BPS equations of $\mathcal{N}=4$ Yang-Mills
on $S^3\times\mathbb{R}$} \label{firstsection}

For convenience, we denote the 3 chiral scalars as $\phi^i$, where $i=1,2,3$ and we
will not use any lowered $SU(3)$ indices. Rather, an $SU(3)$ index on a
conjugate field, $\bar{\phi}^i$, will be understood as anti-fundamental. We begin by
considering the theory on $S^3\times\mathbb{R}$.

The bosonic part of the action of $\mathcal{N}=4$ Yang-Mills theory
on $S^3\times\mathbb{R}$ is
\begin{equation}\label{lagrangian}\begin{split}
  S=\int dt d^3\Omega\tr\left(
  -\frac{1}{4}F_{\mu\nu}F^{\mu\nu}-\frac{1}{2}D_\mu \phi^i D^\mu
  \bar{\phi}^i-\frac{1}{2}|\phi^i|^2+{1\over 4}[\phi^i,\phi^j][\bar{\phi}^i,\bar{\phi}^j]-{1\over 8}[\phi^i,\bar{\phi}^i]^2
  \right),
\end{split}
\end{equation}
and the energy density is
\begin{eqnarray}\label{firstenergydensity}
  \mathcal{E}&=&\tr\left[\frac{1}{2}(F_{03})^2+\frac{1}{2}(F_{12})^2+
  \frac{1}{2}\left(\frac{}{}(F_{a0})^2+(F_{a3})^2\right)\right.\\
  &&\hspace{0.6cm}+\frac{1}{2}\left(\frac{}{}|D_0 \phi^i|^2+|D_3 \phi^i|^2\right)+
  \frac{1}{2}\left(\frac{}{}|D_1 \phi^i|^2+|D_2 \phi^i|^2\right)+
  \frac{1}{2}|\phi^i|^2\\
  &&\hspace{0.6cm}\left.-\frac{1}{4}[\phi^i,\phi^j][\bar{\phi}^i,\bar{\phi}^j]+
  \frac{1}{8}[\phi^i,\bar{\phi}^i]^2\right].
\end{eqnarray}
Here $i=1,2,3$, and the $a=1,2$ and $3$ subscripts label a local
orthonomal frame on $S^3$, where the dreibein may be chosen to be
proportional to the left-invariant 1-forms
\begin{equation}\label{leftinv}\begin{split}
e^1&=\sin\psi d\theta-\cos\psi\sin\theta d\phi \\
e^2&=\cos\psi d\theta+\sin\psi\sin\theta d\phi \\
e^3&=d\psi+\cos\theta d\phi\ .
\end{split}\end{equation}
We define $\phi^i\equiv X^i+iY^i$ with $X^i,Y^i$
hermitian\footnote{We note that the scalar potential of $\mathcal{N}=4$ Yang-Mills can
be decomposed into the F- and D- term potentials as
\begin{equation}
  -\frac{1}{4}{\rm tr}\left(\frac{}{}[X^i,X^i]^2+2[X^i,Y^i]^2+[Y^i,Y^i]^2\right)
  =-\frac{1}{4}{\rm tr}[\phi^i,\phi^j][\bar{\phi}^i,\bar{\phi}^j]+
  \frac{1}{8}{\rm tr}[\phi^i,\bar{\phi}^i]^2\nonumber\ ,
\end{equation}
which we used in the expressions (\ref{lagrangian}) and (\ref{firstenergydensity}) for the Lagrangian and energy
density.} and by a
suitable choice of the time coordinate, we take the scalar mass to
be $1$.

It is shown in appendix \ref{partsappendix} that the energy density
can be written as:
\begin{eqnarray}\label{energythree}
  \mathcal{E}&=&\tr\left[
  \frac{1}{2}\right.\left(F_{12}+\frac{1}{2}[\phi^i,\bar{\phi}^i]\right)^2
  +\frac{1}{2}\left(\frac{}{}F_{a0}- F_{a3}\right)^2
  +\frac{1}{2}\left|\frac{}{}D_0\phi^i- D_3 \phi^i+ i\phi^i\right|^2
  \nonumber\\
  &&\hspace{0.6cm}+\frac{1}{2}\left|\frac{}{}(D_1+iD_2)\phi^i\right|^2
  +\frac{1}{2}(F_{03})^2+\frac{1}{4}\left|\frac{}{}[\phi^i,\phi^j]\right|^2\\
  &&\hspace{0.6cm}\left.
  +\left( F_{a0}F_{a3}+\frac{1}{2}\left(D_0\phi^iD_3\bar{\phi}^i
  +D_0\bar{\phi}^iD_3\phi^i\right)\right)+\frac{i}{2}\left(\bar{\phi}^iD_0
  \phi^i-\phi^iD_0\bar{\phi}^i\right)\right]\ .\nonumber
\end{eqnarray}
The last line of equation (\ref{energythree}) is a sum
of the conserved charges of this theory:
\begin{equation}
  J\equiv\frac{1}{2}\tr\left[F_{a0}F_{a3}+\frac{1}{2}
  \left(D_0\phi^iD_3\bar{\phi}^i+D_0\bar{\phi}^iD_3\phi^i\right)\right]
\end{equation}
is a component of the $SO(4)$ the angular momentum conjugate to
$\psi$, and
\begin{equation}
  Q_i\equiv\frac{i}{2}\tr\left(\bar{\phi}^iD_0
  \phi^i-\phi^iD_0\bar{\phi}^i\right)\ \ \ \ ({\rm no\ sum\ over}\ i)
\end{equation}
are the three $U(1)^3\subset SO(6)$ R-charges. For given values of
these charges, the energy is minimized when the complete-squared
terms on the first and second lines of (\ref{energythree}) vanish.
The Bogomolnyi equations obtained in this way are
\begin{equation}
  F_{12}+\frac{1}{2}[\phi^i,\bar{\phi}^i]=0\ ,\ \
  F_{0a}=F_{3a}\ ,\ \ F_{03}=0
\end{equation}
and
\begin{equation}
  [\phi^i,\phi^j]=0\ ,\ \ D_0\phi^i-D_3 \phi^i+i\phi^i=0\ ,\ \
  (D_1+iD_2)\phi^i=0\ .
\end{equation}
 The energy of a configuration satisfying these equations is
\begin{equation}
  \mathcal{E}=2J+\sum_{i=1}^3 Q_i\ .
\end{equation}
This is a classical version of equation (\ref{bpsrelation}) so that configurations
 satisfying the Bogomolnyi equations above preserve the supersymmetry generated by a single
supercharge and its Hermitian conjugate.

Apart from the above set of BPS equations, we should also impose the
Gauss law constraint to ensure the configuration solves all the equations of
motion. The Gauss law constraint
\begin{equation}
  D^\mu F_{\mu 0}+\frac{i}{2}\left([\phi^i,D_0\bar{\phi}^i]+
  [\bar{\phi}^i,D_0 \phi^i]\frac{}{}\right)=0
\end{equation}
can be rewritten using some of the BPS equations as
\begin{equation}
  D^aF_{a3}+\frac{i}{2}\left([D_3\phi^i,\bar{\phi}^i]-[\phi^i,D_3\bar{\phi}^i]
  \frac{}{}\right)+[\phi^i,\bar{\phi}^i]=0\ .
\end{equation}
Here the covariant derivative $D_a$ contains spin connection as well
as the Yang-Mills connection.

\subsection{Reformulation: Configurations in $\mathbb{R}^4$}
\label{reform}

The $\mathcal{N}=4$ Yang-Mills theory $S^3\times \mathbb{R}$ can be
mapped to the theory defined on $\mathbb{R}^4$ by a conformal
tansformation which places time along the radial direction of
$\mathbb{R}^4$. It will be convenient to consider
the BPS equations that we derived in the this framework. In this
subsection, after reviewing the map between classical configurations
in two theories, we rewrite the BPS equations for configurations in
$\mathbb{R}^4$.

We first introduce $\tau=it$, where $t$ is the
$S^3\times\mathbb{R}$ time coordinate. This changes the positive
frequency modes $e^{-it}$ into $e^{-\tau}$. Then we identify the
radial direction, $r$, of $\mathbb{R}^4$ with $\tau$ through
$r\equiv e^\tau$. The Lagrangian on $S^3\times\mathbb{R}$ should
change to that on $\mathbb{R}^4$ ($iS=-S_E$) where the fields are
related by
\begin{equation}\label{conffields}
  [\phi^i]_{S^3}=r[\phi^i]_{R^4}\ ,\ \ [A_m]_{S^3}=[A_m]_{R^4}\ \
  (m=1,2,3)\ ,\ \ [A_0]_{S^3}=i[A_\tau]_{S^3}
  =ir[A_r]_{R^4}\ .
\end{equation}
Note that we are \textit{not} considering the analytically continued
theory: Euclidean notations are used since it simplifies the
analysis, but we are still considering the \textit{real-time}
physics of this theory by regarding $\tau$ and $A_\tau$ as
imaginary.

We introduce complex coordinates $(Z^1,Z^2)$ which are related to
the spherical coordinates, $(r,\theta,\psi,\phi)$, of equation
(\ref{leftinv}) as
\begin{equation}\begin{split}
Z^1&=r \cos\zeta\ e^{i{\psi+\phi\over 2}}\\
Z^2&=r \sin\zeta\ e^{i{\psi-\phi\over 2}}\ ,
\end{split}\end{equation}
where $\zeta=\theta/2$. The relation between derivatives in these
two coordinate systems can be found in appendix \ref{coordapp}.
Written in the new coordinates which cover $\mathbb{R}^4$, the BPS
equations become:
\begin{eqnarray}
  &&F_{\bar{1}\bar{2}}=0\ ,\ \ F_{I\bar{J}}\bar{Z}^J=0\ ,\ \
  F_{1\bar{1}}+F_{2\bar{2}}+\frac{i}{4}[\phi^i,\bar{\phi}^i]=0\label{bpsgauger4}\\
  &&D_{\bar{I}}\phi^i=0\ ,\ \ [\phi^i,\phi^j]=0\label{bpsscalarbpsr4}
\end{eqnarray}
where $I=1,2$. We note the distinctive fact that complex conjugation
on $S^3\times\mathbb{R}$ becomes complex conjugation plus radial
inversion on $\mathbb{R}^4$ because of the relation $r\equiv e^\tau$
and of the fact that $\tau$ is imaginary. Therefore, when complex
conjugating in $\mathbb{R}^4$, we should simultaneously perform a
coordinate inversion $x^{\prime\mu}=\frac{x^\mu}{r^2}$. With this
in mind, we define a new conjugation operation as $[f(Z^I)]^\star\equiv f^\ast(\bar{Z}_I/r^2)$.
The gauge field transforms under inversion like a derivative, $\partial_\mu=\frac{1}{r^2}\partial^\prime_\mu-\frac{2x_\mu
  x^\nu}{r^4}\partial^\prime_\nu$, so the reality constraint,
$A_\mu=A^{\ast}_\mu$, of the gauge field in $S^3\times\mathbb{R}$
is modified to:
\begin{equation}\begin{split}
  A_\mu&=\frac{1}{r^2}A^\star_\mu-\frac{2x_\mu
  x^\nu}{r^4}A^\star_\nu\ ,\\
  F_{\mu\nu}&=\frac{1}{r^4}F^\star_{\mu\nu}-\frac{2}{r^6}\left(
  x_\mu x^\rho F^\star_{\rho\nu}+x_\nu x^\rho
  F^\star_{\mu\rho}\right)\ ,
\end{split}\end{equation}
where $F^\star_{\mu\nu}=\partial_\mu^\prime
A^\star_\nu-\partial_\nu^\prime A^\star_\mu$. Applying the BPS
equations, we can write the complex conjugation of the scalar and
field strength in $\mathbb{R}^4$ as:
\begin{eqnarray}
  \bar\phi^i&=&\frac{1}{r^2}(\phi^i)^\star\\
  F_{I\bar{J}}&=&\frac{\bar{Z}^I(\epsilon_{\bar{J}\bar{K}}\bar{Z}^K)}
  {r^6}(F_{12})^\star-\frac{i}{4}\frac{(\epsilon_{IK}Z^K)
  (\epsilon_{\bar{J}\bar{L}}\bar{Z}^L)}{r^2}[\phi^i,\bar{\phi^i}]\ .\label{fijconj}
\end{eqnarray}

The equation (\ref{bpsgauger4}) relate most of the components of the field strength
so that the only independent components of the field strength in a BPS configuration
may be taken to be $F_{12}$ and $F_{1\bar1}$. These components are further related by the
reality constraint (\ref{fijconj}).

The details involved in obtaining the Gauss' law constraint in $\mathbb{R}^4$ are
relegated to appendix \ref{gaussapp} and we list only the final constraint here.
Defining curly derivatives as
\begin{equation}
  \mathcal{D}\equiv\frac{1}{r^2}\left(\bar{Z}^2D_1-\bar{Z}^1D_2\right)
   \ ,\ \ \bar{\mathcal{D}}\equiv r^2
   \left(Z^2D_{\bar{1}}-Z^1D_{\bar{2}}\right)\ ,
\end{equation}
the Gauss law constraint is:
\begin{equation}
  \bar{\mathcal{D}}F_{12}+\frac{i}{4}
  [\phi^i,Z\cdot D{\phi^i}^\star]-\frac{i}{4}[\phi^i,{\phi^i}^\star]=0\ .
\end{equation}

\subsection{Axial Gauge}
\label{axialg}

In this section, we will make a convenient choice of gauge which solves some of
the BPS relations and reduces the number of constraints to be
considered. The boundary conditions appropriate for fields in the
radial quantization will also play role in constraining the BPS
solutions.

We make the following choice of gauge:
\begin{equation}
  \bar{Z}^IA_{\bar{I}}=0\ .
\end{equation}
With this choice of gauge, we find several simplifications. First,
the condition $F_{I\bar{J}}\bar{Z}^J=0$ becomes
\begin{equation}
  \bar{Z}\cdot\bar{\partial}A_I=0\ ,
\end{equation}
which says that $A_I$ should be degree $0$ in $\bar{Z}$. We will
restrict our interest to the configurations $A_I$ admitting radial
quantization, namely, those having poles only at $0$ or $\infty$
corresponding to $t=\pm\infty$. Then, for $A_I$ to be of anti-holomorphic degree $0$,
$A_I$ must be a power series in $Z^I$ and $\frac{\bar{Z}^I}{r^2}$.
Furthermore, one finds that the condition $F_{\bar{1}\bar{2}}=0$
becomes linear in $A_{\bar{I}}$ because our gauge condition
$\bar{Z}^IA_{\bar{I}}=0$ implies that $A_{\bar{1}}$ and
$A_{\bar{2}}$ are proportional to each other as matrices so that
$[A_{\bar{1}}, A_{\bar{2}}]=0$. The resulting linear condition
implies
\begin{equation}
  \partial_{\bar{1}}A_{\bar{2}}-\partial_{\bar{2}}A_{\bar{1}}=0
  \ \rightarrow\ \ A_{\bar{I}}=\partial_{\bar{I}}v
\end{equation}
for some matrix $v$ and the gauge condition now becomes
\begin{equation}
  \bar{Z}\cdot\bar{\partial}v=0\ .
\end{equation}
This means that $v$ is degree $0$ in $\bar{Z}$, so that $A_{\bar{I}}$ is
degree $-1$ in $\bar{Z}$. Again allowing $A_{\bar{I}}$ to have poles
only at $0$ and $\infty$, we find that $A_{\bar{I}}$ is
$\frac{1}{r^2}$ times a series expansion of $Z^I$ and
$\frac{\bar{Z}^I}{r^2}$.\footnote{This does not necessarily mean
that the function $v$ should also be a series expansion of $Z^I$ and
$\frac{\bar{Z}^I}{r^2}$.}

Putting together the above observations and the gauge condition
$\bar{Z}^IA_{\bar{I}}=0$, the potential $A_{\bar{I}}$ takes the form
\begin{equation}\label{generalAone}
  A_{\bar{I}}=\frac{\epsilon_{\bar{I}\bar{J}}\bar{Z}^J}{r^4}
  f^\star\left(\frac{\bar{Z}}{r^2},Z\right)\ ,
\end{equation}
where $f^\star$ is an arbitrary function taking the form of series
expansion of the arguments. The most general form of $A_I$,
compatible with the degree constraint and also with
(\ref{generalAone}) through complex conjugation, is given by
\begin{equation}\label{generalA}
  A_I=i\frac{\bar{Z}^I}{r^2}g\left(Z,\frac{\bar{Z}}{r^2}\right)+
  \epsilon_{IJ}Z^J f\left(Z,\frac{\bar{Z}}{r^2}\right)\ ,
\end{equation}
where $g$ is an arbitrary Hermitian matrix function with respect to
the $\star$ operation.

To summarize, we have chosen a gauge, solved
$F_{I\bar{J}}\bar{Z}^J=0$, and expressed all components of the gauge field in terms
of a function $f$ and a Hermitian function $g$. The remaining
equations to be solved are
\begin{equation}\label{BPSeqns}
  F_{1\bar{1}}+F_{2\bar{2}}+\frac{i}{4}[\phi^i,\bar{\phi}^i]=0\ ,\ \
  D_{\bar{I}}\phi^i=0\ ,\ \ [\phi^i,\phi^j]=0
\end{equation}
and the Gauss' Law
\begin{equation}\label{gausseqn}
  \bar{\mathcal{D}}F_{12}+\frac{i}{4}
  [\phi^i,Z\cdot D{\phi^i}^\star]-\frac{i}{4}[\phi^i,{\phi^i}^\star]=0\ .
\end{equation}
These two equations are nonlinear differential equations of the
functions $f$, $g$ and $\phi^i$, which we expect to be difficult to
solve in general. Nevertheless, a class of exact solutions to
these equations can be obtained by imposing additional symmetry
requirements. These solutions are described in appendix \ref{exactsolutions}.

In the next section, we try to analyze the
equations (\ref{BPSeqns}) and (\ref{gausseqn})
approximately in the weakly-coupled regime, in which the
functions are expanded into power series of $g_{YM}$.

\section{Classical $\frac{1}{16}$-BPS Configurations in Weakly
Interacting Theory}
\label{classicalconfigs}

In this section we analyze the differential conditions for the
supersymmetric configurations perturbatively in the weakly-coupled
theory. For the sake of convenience, we first consider the sector
with only gauge fields in section \ref{gaugeonly}, and then
generalize to configuration involving nonzero scalars in section \ref{scalarpert}.

\subsection{Perturbative Expansion in $g_{YM}$ with Scalars set to Zero}
\label{gaugeonly}

We will now use our gauge choice and write out the BPS equations
(\ref{BPSeqns}) and Gauss law (\ref{gausseqn}) in terms of the 2
functions $g(Z,\bar{Z}/r^2)$ and $f(Z,\bar{Z}/r^2)$ which appear in the
gauge potential in equations (\ref{generalAone}) and
(\ref{generalA}). Then we will expand $f$ and $g$ in terms of the
coupling constant $g_{YM}$. We define differential operators as
\begin{eqnarray}
  &&\delta\equiv \frac{1}{r^2}\epsilon^{IJ}\bar{Z}^J\partial_I\ ,\ \
  \bar{\delta}\equiv r^2\epsilon^{\bar{I}\bar{J}}Z^J\partial_{\bar{I}}\ ,\\
  &&\delta\bar{\delta}=r^2\partial_I\partial_{\bar{I}}
  -(\bar{Z}\cdot\bar{\partial})-(Z\cdot\partial)
  (\bar{Z}\cdot\bar{\partial})\ ,\label{pseudolaplacian}\\
  &&\bar{\delta}\delta=r^2\partial_I\partial_{\bar{I}}
  -(Z\cdot\partial)-(Z\cdot\partial)
  (\bar{Z}\cdot\bar{\partial})\ .
\end{eqnarray}
For simplicity, we use this section to record the BPS equations and Gauss law for
configurations where the scalars are turned off and illustrate the
perturbative expansion in that context. Writing the BPS and Gauss
law equations (\ref{BPSeqns}) and (\ref{gausseqn}) in terms of the
functions $f,g$ gives:
\begin{equation}\begin{split}
  0=&g+[f,f^\star]+i\left(\delta f^\star-\bar{\delta}f\right)\\
  0=&-2\bar{\delta}f-\bar{\delta}(Z\cdot\partial)f+
  i\bar{\delta}\delta g+\bar{\delta}[f,g]\\
  &\ \ \ \ -2i[f,f^\star]+i[f^\star,Z\cdot\partial f]
  +[f^\star,\delta g]-i[f^\star,[f,g]]\ .
\end{split}\end{equation}
One can explicitly check, by using the first equation, that the
second equation is purely imaginary. We now treat the Yang-Mills
coupling constant as small expansion parameter. To restore $g_{YM}$
in the above equations which are all omitted, it suffices to put one
$g_{YM}$ in front of each commutator. The functions $f$ and $g$ are
expanded as
\begin{equation}
  f=f_0+g_{YM}f_1+(g_{YM})^2f_2\cdots\ ,\ \
  g=g_0+g_{YM}g_1+(g_{YM})^2g_2\cdots\ .
\end{equation}
The gauge transformation acts as
\begin{equation}
  f\rightarrow U^\star fU+\frac{i}{g_{YM}}U^\star\delta U\ ,\ \
  g\rightarrow U^\star gU+\frac{1}{g_{YM}}U^\star(Z\cdot\partial)U\ ,
\end{equation}
where the unitary matrix $U$ can be written as
\begin{equation}
  U=e^{-ig_{YM}u}\ ,\ \
  u=u_0+g_{YM}u_1+(g_{YM})^2u_2\cdots\ .
\end{equation}
The solution of the free theory is:
\begin{equation}
  f=f_0(Z)+\mathcal{O}(g_{YM})\ ,\ \ g=0+\mathcal{O}(g_{YM})\ ,
\end{equation}
where $f_0(Z)$ is an $N\times N$ matrix whose components are
holomorphic functions of $Z^{1,2}$. The free BPS solution is
therefore given by $N^2$ holomorphic functions.

The differential equations at the next order $\mathcal{O}(g_{YM})$
are:
\begin{eqnarray}\label{pertone}
  g_1&=&[f_0^\star,f_0]+i\left(\bar\delta f_1-\delta
  f_1^\star\right)\ , \\
  F_{12}^{(1)}&=&-(2+Z\cdot\partial)f_1+i[f_0^\star,\delta f_0]
  +\delta\delta f_1^\star-\delta\bar\delta f_1\ .
\end{eqnarray}
Taking the first equation in (\ref{pertone}) to solve for $g_1$, we
write out the $\mathcal{O}(g_{YM})$ part of the second equation as a
condition on $f_1$ and $f_1^\star$:
\begin{equation}\label{linearorder}
  \delta\bar\delta\bar\delta f_1-\bar\delta\delta\delta f_1^\star
  =i\left[\bar\delta f_0^\star,\delta f_0\right]+
  2i\left[f_0^\star,f_0\right]\ .
\end{equation}
In the free theory, the right hand side
of (\ref{linearorder}) vanishes because $g_{YM}=0$ and the zeroth order solution $f_0(Z)$
(with $g_0=0$) solves the BPS equations exactly. We shall explain
below that not all solutions of the free BPS equations can be perturbed to BPS
solutions of (\ref{linearorder}) at nonzero coupling. In fact, for general $f_0(Z)$, we will show that
there can be obstructions to the existence of solutions, $f_1$, to
(\ref{linearorder}). For such obstructions to be absent, $f_0(Z)$
will have to satisfy a set of constraints which we list below.

Arguments in appendix \ref{holomorphyconstraints} show that the
integrability constraint on the right hand side of
equation (\ref{linearorder}) is that it may contain all terms,
$(Z^1)^a(Z^2)^b\left({\bar{Z}^1\over r^2}\right)^c\left({\bar{Z}^2\over r^2}\right)^d$
except those where $a=b=0$ or $c=d=0$. That is $\delta\bar\delta h$ (where $h = \bar\delta f_1$)
should not contain any
holomorphic (or their $\star$ conjugate) terms.

To process this constraint,
we must collect all purely holomorphic terms which may appear in:
\begin{equation}\label{sourceexpress}
\big[ \bar{\delta}f_0^\star,\delta f_0 \big]+2\big[ f_0^\star,f_0
\big]\ .
\end{equation}
We do this by expanding the zeroth order solution $f_0(Z)$ as
\begin{equation}\label{expfo}
  f_0(Z)=\sum_{n_1,n_2=0}^\infty a_{n_1n_2}Y_{n_1n_2}(Z^1,Z^2)\ ,
\end{equation}
where the coefficients $a_{n_1n_2}$ are matrix-valued, and the
functions
\begin{equation}
  Y_{n_1n_2}\equiv\sqrt{\frac{(n_1+n_2+1)!}{2\pi^2(n_1)!(n_2)!}}
  \ (Z^1)^{n_1}(Z^2)^{n_2}\equiv C_{n_1n_2}(Z^1)^{n_1}(Z^2)^{n_2}
\end{equation}
are normalized on unit $S^3$ (i.e., $r=1$) as
\begin{equation}
  \left.\int_{0}^{\frac{\pi}{2}}d\zeta\int_{0}^{2\pi}d\phi
  \int_{0}^{2\pi}d\psi\ (Y_{m_1m_2})^\star Y_{n_1n_2}
  \right|_{r=1}=\delta_{m_1n_1}\delta_{m_2n_2}\ .
\end{equation}
The full holomorphic contribution to the right hand side of equation
(\ref{linearorder}) is calculated in appendix \ref{integoneapp}:
\begin{eqnarray}\label{gaugeonlyconstraint}
  &&\hspace{-1cm}\sum_{n_1,n_2=0}^\infty Y_{n_1n_2}(Z^1,Z^2)
  \sum_{k_1,k_2=0}^\infty\left(k_1\!+\!k_2\!+\!2\right)
  \frac{C_{n_1n_2}C_{k_1k_2}}{C_{n_1\!+\!k_1,n_2\!+\!k_2}}
  \left[a_{k_1k_2}^\ast,\frac{}{}a_{n_1\!+\!k_1,n_2\!+\!k_2}\right]
  \ ,
\end{eqnarray}
where $C_{k_1k_2}=\sqrt{\frac{(k_1+k_2+1)!}{2\pi^2(k_1)!(k_2)!}}$.
The BPS constraints at first order in $g_{YM}$ are therefore given
by setting the coefficients of all independent terms in
(\ref{gaugeonlyconstraint}) to zero. Thus the holomorphic
constraints are parameterized by a pair of non-negative integers
$(n_1,n_2)$:
\begin{equation}
  Q_{n_1n_2}\equiv\sum_{k_1,k_2=0}^\infty\left(k_1\!+\!k_2\!+\!2\right)
  \frac{C_{n_1n_2}C_{k_1k_2}}{C_{n_1\!+\!k_1,n_2\!+\!k_2}}
  \left[a_{k_1k_2}^\ast,\frac{}{}a_{n_1\!+\!k_1,n_2\!+\!k_2}\right]=0\ .
\end{equation}
Since the expression (\ref{sourceexpress}) is explicitly
self-adjoint under the $^\star$ operation, the coefficients of the
purely antiholomorphic terms $\left({\bar{Z}^1\over
r^2}\right)^{n_1}\left({\bar{Z}^2\over r^2}\right)^{n_2}$ are simply
the hermitian conjugates $(Q_{n_1n_2})^\dag$.

We can interpret one of these constraints as the traditional Gauss
law, which arises from the fact that the operator $\delta\bar\delta$
is a Laplacian:
\begin{equation}
  \delta\bar\delta=r^2\partial_I\partial_{\bar{I}}=
  \frac{1}{r}\partial_r r^3 \partial_r+\nabla_{S^3}\ ,
\end{equation}
and integration of the left hand side of (\ref{linearorder}) over a
3-sphere, with fixed $r$, vanishes. The $S^3$ part of the Laplacian
obviously vanishes on integration because it is a total derivative. For the
remaining radial derivative part, we recall that all functions
appearing in our equations are degree $0$ in $\bar{Z}$, which
according to equation (\ref{anglap}) means that
\begin{equation}
  2\bar{Z}\cdot\bar\partial=r\partial_r-i\partial_\psi\ .
\end{equation}
Since the radial derivative part of $\delta\bar\delta$ acts on a degree
$0$ function, we may make the replacement
$\partial_r\stackrel{eff}{=}\frac{i}{r}\partial_\psi$, which makes it
clear that the radial part of the Laplacian is also a total
derivative. We then arrive at the Gauss Law consistency condition:
\begin{equation}
  \int_{S^3}\left(\left[\bar\delta f_0^\star,\delta f_0\right]+
  2\left[f_0^\star,f_0\right]\ \right)=0\ ,
\end{equation}
which we recognize as:
\begin{equation}
  Q_{00}=C_{00}\sum_{k_1,k_2=0}^\infty
  \left(k_1\!+\!k_2\!+\!2\right)
  \left[a_{k_1k_2}^\ast,\frac{}{}a_{k_1k_2}\right]\ .
\end{equation}

To summarize, expanding $f_0$ in spherical harmonics
$f_0=\sum_{n_1,n_2=0}^\infty a_{n_1n_2}Y_{n_1n_2}(Z^1,Z^2)$, $f_0$
can be perturbed to a nearby BPS solution of a weakly interacting
theory only if it solves the equations $Q_{n_1n_2}=0$ for all
nonnegative $n_1,n_2$. We may consider this set of constraints to be
a generalization of the Gauss law.

\subsection{Perturbative Expansion Including Scalars}
\label{scalarpert} Now we generalize the perturbative expansion of
section \ref{gaugeonly} to include the scalar fields. The
supersymmetric configurations of the free theory are parameterized
by four unconstrained holomorphic functions: $f_0$ for the gauge
field, and $\phi^i_0$ for three chiral scalars. Recall that the
zeroth order value of the function $g$ in the gauge potential,
(\ref{generalA}), is zero.

We now turn to the $\mathcal{O}(g_{YM})$ analysis. We write the
following set of unsolved BPS equations
\begin{eqnarray}
  &&F_{1\bar{1}}+F_{2\bar{2}}+\frac{i}{4}[\phi^i,\bar{\phi}^i]=0\\
  &&\bar{\mathcal{D}}F_{12}+\frac{i}{4}[\phi^i,Z\cdot D{\phi^i}^\star]
  -\frac{i}{4}[\phi^i,{\phi^i}^\star]=0\\
  &&D_{\bar{I}}\phi^i=0\ ,\ \ [\phi^i,\phi^j]=0
\end{eqnarray}
in terms of the functions $f$, $g$, $\phi^i$ as in section
\ref{gaugeonly} and expand them in $g_{YM}$. The BPS and Gauss
equations at order $\mathcal{O}(g_{YM})$ give equations for $g_1$,
$f_1$ and $\phi^i_1$ with source terms given by $f_0$, $\phi^i_0$.
The equations are
\begin{eqnarray}\label{perttwofirst}
  g_1&=&i(\bar\delta f_1-\delta f_1^\star)-[f_0,f_0^\star]+
  \frac{1}{4}[\phi_0^i,{\phi^i_0}^\star]\\ \label{perttwosecond}
  \delta\bar\delta\bar\delta f_1-\bar\delta\delta\delta f_1^\star
  &=&i\left([\bar\delta f_0^\star,\delta f_0]+2[f_0^\star,f_0]
  \frac{}{}\right)\\
  &&\frac{i}{4}\left([{\phi^i_0}^\star,\phi^i_0]+
  [\bar\delta{\phi^i_0}^\star,\delta\phi^i_0]+
  (\delta\bar\delta+\bar\delta\delta)
  [\phi^i_0,{\phi^i_0}^\star]
  \frac{}{}\right)\nonumber\\ \label{perttwothird}
  \bar\delta\phi^i_1&=&+i[f_0^\star,\phi^i_0]\\ \label{perttwofourth}
  0&=&[\phi^i_0,\phi^j_0]\ .
\end{eqnarray}
In equations (\ref{perttwofirst}) and (\ref{perttwosecond}), there
is an implicit sum over $i$ on the right hand sides. The first
equation specifies $g_1$ explicitly in terms of other fields, while
the latter three equations require integrability conditions which
generalize the constraints of section \ref{gaugeonly}.

We will now obtain explicit expressions for the constraints. One
class of constraints arises from equation (\ref{perttwosecond}) and
 generalizes the constraints we found in section \ref{gaugeonly}. We will
 label these constraints as $L_{mn}$ and the computation leading to their explicit
 expression may be found in appendix \ref{scalarapp}. We now expand $f_0$ as in equation
 (\ref{expfo}) and the scalars as $\phi^i=\sum_{n_1,n_2=0}^\infty b^i_{n_1n_2} Y_{n_1n_2}$.
The result is that the holomorphic part of the right hand side of
equation (\ref{perttwosecond}) is:
\begin{eqnarray}
  &&\hspace{-1.5cm}\sum_{n_1,n_2=0}^\infty
  Y_{n_1n_2}\sum_{k_1,k_2=0}^\infty
  c^{n_1n_2}_{k_1k_2}\left(
  \left(k_1\!+\!k_2\!+\!2\right)
  \left[a_{k_1k_2}^\ast,\frac{}{}a_{n_1\!+\!k_1,n_2\!+\!k_2}\right]
  +\frac{1}{4}\left(n_1\!+\!n_2\!+\!k_1\!+\!k_2\!+\!1\right)
  \left[b_{k_1k_2}^{i\ast},\frac{}{}b^i_{n_1\!+\!k_1,n_2\!+\!k_2}\right]\right)
  \nonumber\\
\end{eqnarray}
where the coefficient $c^{n_1n_2}_{k_1k_2}$ is defined following the
notation of section \ref{gaugeonly} as:
\begin{equation}
c^{n_1n_2}_{k_1k_2}\equiv {C_{n_1, n_2}C_{k_1, k_2}\over
C_{n_1+k_1,n_2+k_2}}\ .
\end{equation}
The constraint arising from equation (\ref{perttwothird}) is that
the anti-holomorphic part of $[f_0^\star,\phi^i_0]$ should be zero:
\begin{eqnarray}
  \sum_{n_1,n_2=0}^\infty Y^\star_{n_1n_2}
  \sum_{k_1,k_2=0}^\infty c^{n_1n_2}_{k_1k_2}
  \left[a_{n_1\!+\!k_1,n_2\!+\!k_2}^\ast,\frac{}{}b^i_{k_1k_2}\right]=0\ .
\end{eqnarray}
The constraints corresponding to the vanishing of holomorphic parts
are therefore
\begin{equation} \hspace{-1.5cm}\begin{split}
  0=J^i_{n_1n_2}&\equiv \sum_{k_1,k_2=0}^\infty c^{n_1n_2}_{k_1k_2}
  \left[a_{n_1\!+\!k_1,n_2\!+\!k_2}^\ast,\frac{}{}b^i_{k_1k_2}\right] \\
  0=L_{n_1n_2}&\equiv
  \sum_{k_1,k_2=0}^\infty c^{n_1n_2}_{k_1k_2}\left(
  \left(k_1\!+\!k_2\!+\!2\right)
  \left[a_{k_1k_2}^\ast,\frac{}{}a_{n_1\!+\!k_1,n_2\!+\!k_2}\right]
  +\frac{1}{4}\left(n_1\!+\!n_2\!+\!k_1\!+\!k_2\!+\!1\right)
  \left[b_{k_1k_2}^{i\ast},\frac{}{}b^i_{n_1\!+\!k_1,n_2\!+\!k_2}\right]\right)
\end{split}\end{equation}
$L_{n_1n_2}=0$ arise from setting the purely holomorphic part of the
source in (\ref{perttwosecond}) to zero, while $J^i_{n_1n_2}=0$
arise from setting the purely anti-holomorphic part of the source in
(\ref{perttwothird}) to zero. Finally, equation
(\ref{perttwofourth}) is itself a constraint which we will call
\begin{equation}
M^{ij}=[\phi_0^i,\phi_0^j]=0\ .
\end{equation}
The BPS solutions of the free theory should satisfy
$L_{n_1n_2}=J^i_{n_1n_2}=0$ and $M^{ij}=0$ to be lifted to a nearby
BPS solutions of the weakly interacting theory.

\section{Quantization of Classically ${1\over 16}$ BPS Solutions}
\label{quantization} We will now consider the quantization of these
classical solutions. The quantization of classically BPS solutions
has been considered before, often in the context of gravitational
solutions \cite{Mandal:2005wv,Grant:2005qc,Maoz:2005nk}, or probe
branes \cite{Beasley:2002xv,Biswas:2006tj,Mandal:2006tk}, and more
recently in the context of conformal field theory
\cite{Bhattacharyya:2007sa}. The space of all classical solutions to
particular equations of motion may be identified with the classical
phase space of the system. Such an identification is possible
because each classical solution is determined by initial data which
correspond to a specification of ``position'' coordinates and
canonically conjugate momentum coordinates at the initial time. This
provides a map between classical solutions of the equations of
motion and points in phase space. We obtained constraints on the
free BPS solutions which, when satisfied, allow the free solutions
to be lifted to solutions at infinitesimal coupling so that a point
in the phase space of the $\mathcal{N}=4$ SYM $1/16$ BPS sector is
identified by specifying the holomorphic functions
$f_0(Z_1,Z_2),\phi^i_0(Z_1,Z_2)$ in such a way that the constraints,
$L_{n_1n_2}=J^i_{n_1n_2}=0$ and $M^{ij}=0$, are all satisfied.

\subsection{Quantization Prescription}
Since the phase space above is described with the use of
constraints, it will be most convenient to perform a constrained
quantization, using the symplectic form of free field theory and
attempting to include the constraints appropriately. Thus, we first consider the
symplectic form of free $\mathcal{N}=4$ Yang-Mills theory, evaluated
on the space of $1/16$ BPS solutions spanned by unconstrained
$f_0(Z)$ and $\phi^i(Z)$.

The contribution of the $U(N)$ gauge field to the symplectic form of
the free theory is given by $N^2$ copies of a $U(1)$ gauge field. A
$U(1)$ gauge field contributes a symplectic form
\begin{equation}\label{gaugesymp}
  \omega=\int_{S^3}dF^{0i}_S\wedge dA_i^S=
  i\int_{S^3}d\left(x\cdot\partial A^\mu\right)\wedge dA_\mu\ ,
\end{equation}
where $A_S^i$ is defined in (\ref{conffields}). 

For the free field theory BPS solution
\begin{equation}
  A_I=\epsilon_{IJ}Z^J f_0\ ,\ \ A_{\bar{I}}=\frac{\epsilon_{\bar{I}\bar{J}}
  \bar{Z}^J}{r^4}f_0^\star\ ,
\end{equation}
one obtains
\begin{eqnarray}
  x\cdot\partial A_I&=&(Z\cdot\partial)A_I=\epsilon_{IJ}Z^J
  \left(Z\cdot\partial+1\right)f_0\nonumber\\
  x\cdot\partial A_{\bar{I}}&=&(Z\cdot\partial-1)A_{\bar{I}}=
  \frac{\epsilon_{\bar{I}\bar{J}}\bar{Z}^J}{r^4}
  (Z\cdot\partial-3)f_0^\star\ .
\end{eqnarray}
We plug this into the symplectic form (\ref{gaugesymp}) and find
that the symplectic form evaluated on the $1/16$ BPS solutions is
\begin{eqnarray}
  \omega&=&2i\int_{S^3}d\left(x\cdot\partial A_I\right)\wedge dA_{\bar{I}}+
  d\left(x\cdot\partial A_{\bar{I}}\right)\wedge dA_I\\
  &=&2i\int_{S^3}4df_0\wedge df_0^\star+d(Z\cdot\partial f_0)\wedge df_0^\star
  -df_0\wedge d(Z\cdot\partial f_0^\star)\ .\nonumber
\end{eqnarray}
Expanding $f_0=\sum_{n_1n_2}a_{n_1n_2}Y_{n_1n_2}$ as in section
\ref{gaugeonly}, one obtains
\begin{equation}
  \omega
  =2i\sum_{n_1n_2}da_{n_1n_2}\wedge da_{n_1n_2}^\ast
  \left(4+2(n_1\!+\!n_2)\frac{}{}\right)\ .
\end{equation}
This means that
\begin{equation}
  A_{n_1n_2}^\ast\equiv 2\sqrt{n_1\!+\!n_2\!+\!2}\ a_{n_1n_2}\
\end{equation}
are normalized creation operators. When quantized, they will satisfy
$[{A_{m_1m_2}}^{a}_{\ b}, ({A_{n_1n_2}}^c_{\ d})^\dag
]=\delta_{m_1,n_1}\delta_{m_2,n_2}\delta^a_c\delta_b^d$ .

The contribution of a scalar field to the symplectic form is
\begin{eqnarray}
  \omega&=&\int_{S^3}d(\partial_0{\phi_S}^\ast)\wedge d\phi_S
  +d(\partial_0\phi_S)\wedge{\phi_S}^\ast\\
  &=&i\int_{S^3}2d\phi\wedge d{\phi}^\star+d(Z\cdot\partial\phi)\wedge
  d\phi^\star-d\phi\wedge d(Z\cdot\partial\phi^\star)\ ,\nonumber
\end{eqnarray}
where $\phi_S$ is a scalar on $S^3\times\mathbb{R}$ and $\phi$ is a
scalar field after a conformal transformation to $\mathbb{R}^4$,
i.e. $\phi_S$ and $\phi$ are related as in equation
(\ref{conffields}). Expanding $\phi=\sum_{n_1,n_2} b_{n_1n_2}
Y_{n_1n_2}$ as in previous sections, one obtains
\begin{equation}
  2i\sum_{n_1,n_2=0}^\infty db_{n_1n_2}\wedge db_{n_1n_2}^{\ast}
  (n_1+n_2+1)\ .
\end{equation}
This implies that
\begin{equation}
  B^{\ast}_{n_1n_2}\equiv\sqrt{2(n_1+n_2+1)}\ b_{n_1n_2}
\end{equation}
are normalized creation operators with standard commutation
relations.

The symplectic form of free $\mathcal{N}=4$ SYM in the bosonic
sector is simply the sum of the contributions from the $N^2$ $U(1)$
gauge fields and the 3 scalars. We will now consider the constraints
of section \ref{scalarpert}, promoting the coefficients
$A_{n_1n_2},A^\ast_{n_1n_2},B_{n_1n_2},B^\ast_{n_1n_2}$ to
creation/destruction operators.

First we will define more convenient creation and annihilation
operators as\footnote{Abusing the notation, we write
$\alpha_{k_1k_2}^\dag$ and $\beta_{k_1k_2}^\dag$ even if they are
not Hermitian conjugates of $\alpha_{k_1k_2}$ and $\beta_{k_1k_2}$
respectively.}:
\begin{equation}\label{classosc}\begin{split}
\alpha_{k_1k_2}&=\sqrt{k_1+k_2+2}\ C_{k_1k_2}\ A_{k_1k_2}\qquad
\alpha^\dagger_{k_1k_2}={1\over C_{k_1k_2} \sqrt{k_1+k_2+2}}A^\ast_{k_1k_2}  \\
\beta_{k_1k_2}&={C_{k_1k_2}\over
\sqrt{k_1+k_2+1}}B_{k_1k_2}\qquad\qquad\quad
\beta^\dagger_{k_1k_2}={\sqrt{k_1+k_2+1}\over
C_{k_1k_2}}B^\ast_{k_1k_2}
\end{split}\end{equation}
These operators are convenient because they have standard
commutation relations and they appear to be more natural variables
for expressing the constraints. We will use a shorthand for the
subscripts as $X_{k_1k_2}\sim X_k$ to make equations more
transparent. In terms of the new variables, the constraints in the
classical theory become operators:
\begin{eqnarray}\label{constraints}
L_n&=&{1\over 2}\sum_{k=0}^\infty
[\alpha_{n+k},\alpha_k^{\dagger}]+[\beta^i_{n+k},\beta^{i\dagger}_k]\\
J^i_n&=&\sum_{k=0}^\infty[\alpha_{n+k},\beta^{i\dagger}_k]
\label{constraints2}\\
M^{ij}_n&=&\sum_{k=0}^n \left({}^{k_1+k_2}C_{k_1}
{}^{n_1+n_2-k_1-k_2}C_{n_1-k_1}\right)[\beta^i_k,\beta^j_{n-k}]
\label{constraints3}
\end{eqnarray}
up to overall multiplicative constants, which we drop. To resolve the
normal-ordering ambiguity that occurs for $L_0$, we define $L_0$ to
be the normal-ordered one, having annihilation operators on the right side
of the creation operators. ${}^{k_1+k_2}C_{k_1}$ is the binomial coefficient
$k_1+k_2$ choose
$k_1$ and we have suppressed the $U(N)$ matrix indices.


Now we must consider how to impose the constraints correctly. Classically, $L_k,L_l^\dag,J^i_m,$ $J^{i\dag}_n,M^{ij}_p,M^{ij\dag}_q$
are all constrained to be zero. We will not perform a
systematic quantization of these first class constraints, rather, we
adopt an approach similar to the Old Covariant Quantization (OCQ).
Since we will find that our prescription results in the same
cohomology as described in section \ref{scalarsector}, we
presume that a rigorous quantization will lead to the same result.

As in OCQ, we will first quantize the $1/16$ BPS sector without
considering the dynamical constraints,
(\ref{constraints})-(\ref{constraints3}), and then impose the
constraints as operator relations in Hilbert space. Explicitly, we
define a vacuum $|0\rangle$ that satisfies
$\alpha_k|0\rangle=\beta^i_k|0\rangle=0$ for all $k$ and let the
$\alpha_k^\dag,\beta^{i\dag}_k$ operate on $|0\rangle$ to produce an
unconstrained Hilbert space. This unconstrained Hilbert space
corresponds simply to the $1/16$ BPS sector of the free
$\mathcal{N}=4$ SYM theory.

Now we impose the constraints in this Hilbert space. Following OCQ
terminology, we will call any state which remains BPS at
infinitesimal coupling ``physical'' and require such a state,
$|\psi\rangle$, to satisfy:
\begin{equation}
L_k|\psi\rangle=J^i_m|\psi\rangle=M^{ij}_p|\psi\rangle=0.
\end{equation}
This ensures that the matrix elements of all the constraints are
zero between any two physical states. For example, if
$|\psi\rangle,|\chi\rangle$ are physical states, then we have:
\begin{equation}
\langle\psi|L_k^\dag|\chi\rangle=\langle  L_k\psi|\chi\rangle=0\ .
\end{equation}

One might wonder whether we should require $L_n$ or $L_n^\dag$ to
annihilate physical states, since either condition would be
sufficient to set all matrix elements of all constraints to zero.
Although being a bit \textit{ad hoc}, it is clear that we should
require $L_n$ and not $L_n^\dag$ to annihilate physical states. This
is because we want states made of symmetrized gauge invariant scalar
zero modes $(\beta_{00}^i)^\dag$, which belong to the $\frac{1}{8}$
BPS Hilbert space, to be physical. If we set $L_n|\Psi\rangle=0$,
where $|\Psi\rangle$ is such a $\frac{1}{8}$ BPS state, then the
only term of $L_n$ which acts nontrivially on $|\Psi\rangle$ is
$\sum^3_{i=1}[(\beta_{00}^i)^\dag,\beta_{00}^i]$ in $L_{00}$, which
simply requires $U(N)$ gauge invariance. If we set
$L_n^\dag|\Psi\rangle=0$, however, terms of the form
$[\beta_{00}^i,(\beta_{n_1n_2}^i)^\dag]$ in $(L_{n_1n_2})^\dag$ act
nontrivially and these $\frac{1}{8}$ BPS states would not be
physical.

As a simple illustration and test of the above prescription, we will
try to identify a single trace physical state made of a single
scalar only. We start from
\begin{equation}
  |2,2\rangle\equiv
  {\rm tr}\left(b_1^\dag b_1^\dag+2b_0^\dag b_2^\dag\right)
  |0\rangle\ ,
\end{equation}
where $b_n^\dag\equiv {\beta^3_{n0}}^\dag$ is a creation operator
for the scalar $\phi^3$, associated with $(Z^1)^n$. This state maps
by the state operator map to\footnote{Note that under the state operator map, the
operator $D_{+\dot{+}}^n\bar\phi^3$ maps to the state
$n!\beta_{n0}^3$} (following the conventions used in section
 \ref{cohomologyapproach}):
\begin{equation}
  (\partial_{+\dot{+}})^2{\rm tr}(\bar\phi^3)^2=
  2{\rm tr}\left(\frac{}{}(D_{+\dot{+}}\bar\phi^3)^2+\bar\phi^3
  D_{+\dot{+}}^2\bar\phi^3\right)\ .
\end{equation}
which according to section \ref{scalarsector} is an allowed $1/16$
BPS state at infinitesimal coupling.

We will now check that $|2,2\rangle$ is annihilated by $L_n$.
It is clear that only the $L_{n0}$ constraints need to be
checked. $L_{n0}|2,2\rangle=0$ for $n\geq 3$ because
the state does not involve a creation $b_n^\dag$ with $n\geq 3$
and $L_{00}|2,2\rangle=0$ is automatic because the trace guarantees
gauge-invariance. Finally, from $[(b_m)^a_{\ b},(b_n^\dag)^c_{\
d}]=\delta_{mn}\delta^a_d\delta^c_b$, we find
\begin{equation}
  (L_{20})^a_{\ b}|2,2\rangle=2\left(\frac{}{}
  (b_0^\dag)^{a}_{\ c}(b_2)^c_{\ b}-(b_0^\dag)^{c}_{\ b}
  (b_2)^{a}_{\ c}\right){\rm tr}(b_0^\dag b_2^\dag)|0\rangle
  =2[b_0^\dag,b_0^\dag]^a_{\ b}|0\rangle=0\ ,
\end{equation}
and
\begin{equation}
  (L_{10})^a_{\ b}|2,2\rangle=
  [b_0^\dag,b_1]^a_{\ b}{\rm tr}(b_1^\dag b_1^\dag)|0\rangle+
  2[b_1^\dag,b_2]^a_{\ b}{\rm tr}(b_0^\dag b_2^\dag)|0\rangle=
  2\left([b_0^\dag,b_1^\dag]+[b_1^\dag,b_0^\dag]\right)|0\rangle
  =0
\end{equation}
So that $|2,2\rangle$ is a physical state if we require that
physical states be annihilated by $L_n$ as discussed above.

When considering the $J^i_m$ constraint, one may again wonder
whether we should require $J^i_m$ or $(J^i_m)^\dag$ to annihilate
physical states. We again use the $\frac{1}{8}$ BPS primaries as a
guide. The $J^i_{n_1n_2}$ operators automatically annihilate the
above $\frac{1}{8}$ BPS primaries because of the gauge field
annihilation operators $\alpha_{n_1\!+\!k_1,n_2\!+\!k_2}$ appearing
in every term of $J^i_{n_1n_2}$, but the $(J^i_{n_1n_2})^\dag$
operators do not annihilate these states. Therefore we must require
that physical states satisfy $J^i_n|\psi\rangle=0$.


It is clear that we must impose $M^{ij}_m|\psi\rangle=0$ because
$(M^{ij}_m)^\dag$ contains only creation operators and therefore
cannot annihilate the $\frac{1}{8}$ BPS primaries. On the other
hand, $M^{ij}_m$ does annihilate these states because it annihilates
states which are symmetrized on the $SU(3)$ index.

Finally, our quantization prescription is that we impose constraints
on the free $1/16$ BPS Hilbert space: States that remain $1/16$ BPS
at infinitesimal coupling, (physical states), should be annihilated
by $L_n,J^i_n,M^{ij}_n$.

\subsection{Relation between Classically Derived Constraints
and the Analysis of Section \ref{scalarsector}}

We will now map the results of the classical analysis to equation (\ref{Q}). We collect the oscillators $\alpha_{k_1k_2},\beta^i_{k_1k_2}$, which resulted from the quantization of the classical solutions (\ref{classosc}), into fields similar to those defined above, which will allow us to consider the BPS constraints obtained in the classical analysis in more detail\footnote{Recall that we are abusing notation: $\alpha_{k_1k_2},\alpha_{k_1k_2}^\dag$ are creation and destruction operators with standard commutation relations, but $\alpha_{k_1k_2}^\dag$ is not really the hermitian conjugate of $\alpha_{k_1k_2}$}:
\begin{equation}\begin{split}
\beta^i(\bar{z})&=\sum_{k_1k_2} {(k_1+k_2+1)!\over k_1!k_2!}\beta^i_{k_1k_2}\bar{z}_1^{k_1}\bar{z}_2^{k_2}\ , \qquad
\beta^{i\dag}(z)=\sum_{k_1k_2}
\beta^{i\dagger}_{k_1k_2}z_1^{k_1}z_2^{k_2}\\
\alpha(\bar{z})&=\sum_{k_1k_2} {(k_1+k_2+1)!\over k_1!k_2!}\alpha_{k_1k_2}\bar{z}_1^{k_1}\bar{z}_2^{k_2}\ , \qquad
\alpha^\dag(z)=\sum_{k_1k_2}
\alpha^\dagger_{k_1k_2}z_1^{k_1}z_2^{k_2}\ ,
\end{split}\end{equation}
We have defined these fields such that
holomorphic fields are creation operators and anti-holomorphic fields are
destruction operators. With these definitions, we can write the constraints obtained in the classical analysis
of section \ref{quantization} as:
\begin{equation}\begin{split}
L_{W_L}&={1\over 2}\int_{S^3}\ Tr\
W_L(z)[\alpha(\bar{z}),\alpha^\dag(z)]
+W_L(z)[\beta^i(\bar{z}),\beta^{i\dag}(z)]\\
J^i_{W_J}&={1\over 2}\int_{S^3}\ Tr\ W_J(z)[\alpha(\bar{z}),\beta^{i\dag}(z)]\\
M^{ij}_{W_M}&={1\over 2}\int_{S^3}\ Tr\ W_M(z) [{1\over
1+\bar{z}\cdot\bar\partial}\beta^i(\bar{z}),{1\over 1+\bar{z}\cdot\bar\partial}\beta^j(\bar{z})]\ ,
\end{split}\end{equation}
where the integrals are over the unit sphere.
The constraints are parameterized by the matrices
$W_L(z),W_J(z),W_M(z)$. We recover the constraints as listed previously by:
\begin{equation}
L_n=L_{z^n}, \quad J^i_n=J^i_{z^n}, \quad M^{ij}_n=M^{ij}_{z^n}\ .
\end{equation}
The notation here means that to recover $L_{n_1n_2}$ for example, we let $W_L=z_1^{n_1}z_2^{n_2}$.
The commutation relations of the local fields are for the scalars, for example:
\begin{equation}
[\beta(\bar{z}),\beta^\dag(w)]=(1+\bar{z}\cdot \bar\partial)
{1\over 1-\bar{z}_1w_1-\bar{z}_2w_2}={1\over (1-\bar{z}_iw_i)^2}
\end{equation}
The right hand side of this commutation relation acts like a delta function
for integration of holomorphic functions on $S^3$. That is:
\begin{equation}\begin{split}
\int_{S^3}  {f(z)\over (1-\bar{z}_iw_i)^2}=\Omega_3f(w)
\end{split}\end{equation}
This is a type of cauchy integral formula in 2 complex dimensions.

Now we list the action of the constraints on the fields
$\alpha^\dag(z),\beta^{i\dag}(z)$.
\begin{equation}\begin{split}
[L_{W_L},(\beta^{i\dag}(z))^{pq}]&=
[\beta^{i\dag}(z),W_L(z)]^{pq}\\
[L_{W_L},(\alpha^\dag(z))^{pq}]&=
[\alpha^\dag(z),W_L(z)]^{pq}\\
[J^i_{W_J},(\alpha^\dag(z))^{pq}]&=
[\beta^{i\dag}(z),W_J(z)]^{pq}\\
\end{split}\end{equation}
the lower $i$ index here is the $SU(3)$ index of the scalar and $p,q$ are
gauge indices.

Further we have:
\begin{equation*}
[{1\over
1+\bar{z}\cdot\bar\partial}(\beta^i(\bar{z}))^{mn},(\beta^{j\dag}(w))^{pq}]
={\delta^{ij}\delta^{mq}\delta^{np}\over
1-\bar{z}^aw^a}
\end{equation*}
Defining $\Phi(\bar{z})={1\over 1+\bar{z}\cdot\bar{\partial}}\beta(\bar{z})$, this gives:
\begin{equation}
[M_{W_M}^{ij},(\beta^\dag_k)^{pq}(w)]=\int_{S^3}
\left\{[\Phi^j(\bar{z}),W_M(z)]^{pq}\delta^{ik}-[\Phi^i(\bar{z}),W_M(z)]^{pq}\delta^{jk}\right\}
{1\over 1-\bar{z}\cdot w}
\end{equation}
The right hand side still contains annihilation operators $\Phi(\bar{z})$
so $M$ naturally acts on pairs of creation fields
$\beta^{k\dag}(z)\beta^{l\dag}(z)$ and we obtain:
\begin{equation}\label{Mconst}\begin{split}
\hspace{-1cm}[M^{ij}_{W_M},(\beta^{k\dag}(w))^{pq}(\beta^{l\dag}(w))^{st}]
&=
(\delta^{ik}\delta^{jl}-\delta^{il}\delta^{jk})\left(W_M^{sq}(w)\delta^{pt}-W_M^{pt}(w)\delta^{sq}\right)\ .
\end{split}\end{equation}
We have assumed here that the operator
$(\beta^{k\dag}(w))^{pq}(\beta^{l\dag}(w))^{st}$ is acting on the
vacuum, that is it stands for
$(\beta^{k\dag}(w))^{pq}(\beta^{l\dag}(w))^{st}|0\rangle$. We
therefore dropped all terms containing destruction operators. The
factor $(\delta^{ik}\delta^{jl}-\delta^{il}\delta^{jk})$ in equation
(\ref{Mconst}) indicates that the $M^{ij}_{W_M}$ constraint requires
$1/16$ BPS operators at infinitesimal coupling to be symmetrized on
the $SU(3)$ index. In fact, similar factors which are antisymmetric
in the $SU(3)$ indices, appear when $M^{ij}_{W_M}$ acts on arbitrary
single-trace states so that $M^{ij}_{W_M}$ requires symmetrization
of $SU(3)$ indices inside traces quite generally.

With these considerations in hand, we now consider the quantization
prescription obtained from the classical analysis of section
\ref{quantization} in cohomological terms. The $L_{W_L}$ constraint
acts on the fields $\alpha^\dag(z),\beta^{i\dag}(z)$ as a gauge
transformation and requires all $1/16$ BPS operators at
infinitesimal coupling to be traces of the fields
$\alpha^\dag(z),\beta^{i\dag}(z)$, where all fields inside a given
trace are at the same position. Therefore it is clear that $L_{W_L}$
implements the same gauge invariance constraint discussed in section
\ref{scalarsector}, with the parameter, $W_L$, of the gauge
transformation mapping to a derivative of the gaugino, $W_L(z)\sim
z^{\dot{\alpha}}(1+z\cdot\partial)^{-1}
\bar{\lambda}_{\dot{\alpha}}(z)$, in the cohomology picture. The
$J_{W_J}$ constraint arises in equation (\ref{Q}) from the action of
$Q^1_-$ on $f(z)$ which produces $[\psi_{n+}(z),\bar{\phi}^n(z)]$.
The parameter $W_J$ maps to the chiralino $W_J(z)\sim \psi_{n+}(z)$
in the cohomology language. Finally, the $M^{ij}_{W_M}$ constraints
correspond to the action of the superconformal generator $S_4^-$ on
commutators of scalars. In section \ref{scalarsector} we worked in
terms of cohomology and this superconformal constraint from $S_4^-$
was interpreted as the exactness condition that arises because
$[\bar{\phi}^n(z),\bar{\phi}^p(z)]$ appears on the right hand side
of (\ref{Q}).

Now it is clear that the analysis of the classical configurations
and their quantization agree with the rules obtained in section
\ref{scalarsector}.

\section{Concluding Remarks}

In this paper we studied the $\frac{1}{16}$-BPS states of the
weakly-coupled $\mathcal{N}=4$ Yang-Mills theory. We formulated the
problem in terms of local fields which generate covariant
derivatives acting on fields. In particular, we thoroughly
investigated the $\frac{1}{16}$-BPS cohomology made of bosonic
letters. We obtained the exact partition function in special limits,
and also an upper bound partition function which turns out to be
useful. We also gave a physical interpretation of the local fields
by studying a set of classical BPS equations for $\frac{1}{16}$-BPS
configurations in the bosonic sector, and suitably quantizing them.

The classical cohomology we consider is in 1-to-1 correspondence
with the zero eigenstates of the 1-loop Hamiltonian acting on the
Hilbert space of $\frac{1}{16}$-BPS states in the free Yang-Mills
theory. In this sense, our analysis generalizes and strengthens the
result of \cite{Janik:2007pm} to the finite $N$ case in certain
subsectors or energy regime.

Using our exact partition function for scalars derived in special
limits, we have also shown that a dual description in the strong
coupling regime, in terms of giant gravitons, sometimes predicts
exactly the same result as ours. We also pointed out that our exact
result shows qualitatively similar behaviors to the result obtained
from a naive quantization of dual giant gravitons. The error of the
latter result is argued to be due to the intersection of multiple
dual giant gravitons. As we commented in section \ref{threefour}, it is
interesting to see if one can correctly quantize them and obtain the
exact partition function.

We used our upper bound partition function to conclusively argue
that the asymptotic degeneracy with large charges is not big enough
to form a supersymmetric black hole.

The most important problem which follows our analysis, combined with
the conjecture that weakly coupled Yang-Mills theory captures the
exact supersymmetric spectrum, should be to find the large number of
$\frac{1}{16}$-BPS states in the high energy regime, with their
entropy scaling like $N^2$ if $E\sim N^2$, and furthermore to
identify it with the entropy of the known supersymmetric black holes
\cite{Kunduri:2006ek}. Since our analysis in the bosonic sector was
quite comprehensive, perhaps fermionic letters should play important
roles. See \cite{Berkooz:2006wc} for some attempts in this
direction.

A more modest question along the similar direction would be to
understand the finite $N$ cohomology that we investigated in this
paper, or any generalization thereof, in the strongly-coupled regime
in terms of giant gravitons. Even if the giant graviton provides an
effective description of BPS states in the energy regime around
$E\sim N$, it hopefully could also give us some new insights or clue
towards understanding the even higher energy regime. For the latter
purpose, one probably would have to go beyond the solutions of
\cite{Kim:2006he} by exciting other degrees of freedom (see
\cite{Sinha:2007ni} also).

\vskip 0.7cm

\hspace*{-0.8cm} {\bf\large Acknowledgements}

\vskip 0.4cm

\hspace*{-0.7cm} We would like to thank Pallab Basu, Amihay Hanany,
Rudra Jena, Subhaneil Lahiri, Ki-Myeong Lee, Sangmin Lee, Sungjay
Lee, Suvrat Raju, and  Toby Wiseman. 
L.G and S.K. also thank the warm hospitality
of TIFR theory group during their visits while this work initiated.
The work of S.M. was supported in part by a Swarnajayanti Fellowship.
S.M. would also like to acknowledge the steady and generous support
of the people of India for research in the basic sciences.


\appendix
\section{Action of Supercharges on the Letters of $\mathcal{N}=4$ Super Yang-Mills}
\label{transformations}

The $SU(4)$ vector $Q^i_\alpha$ with $i=1,\ldots,4$ has been divided
into a special supercharge $Q_\alpha$ and an $SU(3)$ vector
$Q^m_\alpha$ with $m=1,\ldots,3$. In terms of the $SU(4)$ notation,
$Q_\alpha$ is the $4$th component and $Q^m_\alpha$ correspond to the
$1,2,3$ components of the $SU(4)$ vector. The indices $m,n,p$ below
run from $1,\ldots,3$ and the indices $\alpha,\beta,\gamma,\delta$
run over $1,2$. The transformation rule of $\mathcal{N}=4$
Yang-Mills theory, say in the appendix of \cite{Biswas:2006tj}, is
decomposed as
\begin{equation}\begin{split}
[Q^m_\alpha,\bar{\phi}^n]&=-\epsilon^{mnp}\psi_{p\alpha}\\
[Q_\alpha,\bar{\phi}^n]&=0\\
[Q^m_\alpha,\phi_p]&=-\delta^m_p \lambda_\alpha\\
[Q_\alpha,\phi_p]&=\psi_{p\alpha}\\
\{Q^m_\alpha,\lambda_\beta\}&=\epsilon_{\alpha\beta}\epsilon^{mnp}[\phi_n,\phi_p]\\
\{Q_\alpha,\lambda_\beta\}&=2if_{\alpha\beta}+\epsilon_{\alpha\beta}[\phi_k,\bar{\phi}^k]\\
\{Q^m_\alpha,\psi_{n\beta}\}&=2i\delta^m_n
f_{\alpha\beta}-2\epsilon_{\alpha\beta}[\bar{\phi}^m,\phi_n]+\epsilon_{\alpha\beta}\delta^m_n[\bar{\phi}^k,\phi_k]\\
\{Q_\alpha,\psi_{n\beta}\}&=-\epsilon_{\alpha\beta}\epsilon_{nmp}[\bar{\phi}^m,\bar{\phi}^p]\\
\{Q^m_\alpha,\bar{\lambda}_{\dot{\beta}}\}&=2iD_{\alpha\dot{\beta}}\bar{\phi}^m\\
\{Q_\alpha,\bar{\lambda}_{\dot{\beta}}\}&=0\\
\{Q^m_\alpha,\bar{\psi}^n_{\dot{\beta}}\}&=-2i\epsilon^{mnp}D_{\alpha\dot{\beta}}\phi_p\\
\{Q_\alpha,\bar{\psi}^n_{\dot{\beta}}\}&=-2iD_{\alpha\dot{\beta}}\bar{\phi}^n\\
[Q^m_\alpha,A_{\beta\dot{\gamma}}]&=-\epsilon_{\alpha\beta}\bar{\psi}^m_{\dot{\gamma}}\\
[Q_\alpha,A_{\beta\dot{\gamma}}]&=-\epsilon_{\alpha\beta}\bar{\lambda}_{\dot{\gamma}}\\
[Q^m_\alpha,f_{\beta\gamma}]&=\epsilon_{\alpha\{\gamma}D_{\beta\}}^{\dot{\delta}}\bar{\psi}_{\dot{\delta}}^m\\
[Q_\alpha,f_{\beta\gamma}]&=\epsilon_{\alpha \{
\gamma}D_{\beta\}}^{\dot{\delta}}\bar{\lambda}_{\dot{\delta}}
=-{i\over 2}\epsilon_{\alpha\gamma}[\bar{\phi}^m,\psi_{m\beta }]-{i\over 2}\epsilon_{\alpha\beta}[\bar{\phi}^m,\psi_{m\gamma }]\\
[Q^m_\alpha,f_{\dot{\beta}\dot{\gamma}}]&=-D_{\alpha\{\dot{\beta}}\bar{\psi}_{\dot{\gamma}\}}^m\\
[Q_\alpha,f_{\dot{\beta}\dot{\gamma}}]&=-D_{\alpha\{\dot{\beta}}\bar{\lambda}_{\dot{\gamma}\}}
\end{split}\end{equation}
In the equation for $[Q_\alpha,f_{\beta\gamma}]$, we have used the
fermionic equation of motion:
\begin{equation}\begin{split}
&D_{\alpha\dot{\beta}}\bar{\lambda}^{\dot{\beta}}=i[\bar{\phi}^m,\psi_{m\alpha}]\\
&D_{\beta\dot{\alpha}}\lambda^\beta =i[\phi_m,\bar{\psi}^m_{\dot{\alpha}}]
\end{split}\end{equation}
The action of the supersymmetries $\bar{Q}^i_{\dot{\alpha}}$ with $i=1,\ldots,4$ are given by
taking the simple complex conjugate of the above relations. With these definitions, the algebra
satisfies $\{Q^i_\alpha,\bar{Q}_{j\dot{\beta}}\}X=2i\delta_j^iD_{\alpha\dot{\beta}}X$ up to a gauge
transformation.

\section{Checks of the $SU(2)$ Partition Function}
\label{checks}

We will now check the $N=2$ partition function of section
\ref{utwopartition} explicitly. It suffices to check the $SU(2)$
gauge group instead of $U(2)$, since the contribution from the
overcall $U(1)$ is always factored out as $Z_1(\mu_i,\theta_a)$ as
is also true in (\ref{utwopartition}). Here we only record the check
for the operators that contain a single instance of a single
derivative. We also checked the operators involving two derivatives
of the same kind (i.e. checks at the next order in angular momentum
charge) from an analysis similar to this section (more involved),
and also find agreement with the partition function of section
\ref{utwopartition}. The $SU(2)$ partition function that we would
like to check is
\begin{equation}\label{b1}
  Z_2(\mu_1,\mu_2,\theta)=\prod_{k=0}^\infty
  \frac{1-\mu_1^2\mu_2^2\theta^{k}}{(1-\mu_1^2\theta^{k})(1-
  \mu_2^2\theta^{k})(1-\mu_1\mu_2\theta^{k})}\ .
\end{equation}
The first two factors in the denominator of the $U(2)$ partition function
(\ref{utwo}) are omitted, since, in $SU(2)$, we do not have the
generators $\partial^k{\rm tr}(X)$ and $\partial^k{\rm tr}(Y)$. The
relation or syzygy \ref{syz} reduces to
\begin{equation}
  {\rm tr}\left(X^2\right){\rm tr}\left(Y^2\right)
  =\left({\rm tr}(XY)\frac{}{}\right)^2\ ,
\end{equation}
and the numerator of the partition function remains unchanged. The Taylor expansion of
the above partition function in terms of $\theta$ is
\begin{eqnarray}
  \hspace*{-2cm}Z_2&=&\frac{1+\mu_1\mu_2}{(1-\mu_1^2)(1-\mu_2^2)}\left\{
  1+\theta\left(\mu_1^2+\mu_2^2+\mu_1\mu_2-\mu_1^2\mu_2^2\right)\frac{}{}
  \right.\\
  &&\left.\frac{}{}+\theta^2\left(\mu_1^4+\mu_2^4+\mu_1^2\mu_2^2
  +\mu_1^3\mu_2+\mu_1\mu_2^3-\mu_1^4\mu_2^2-\mu_1^2\mu_2^4-\mu_1^3\mu_2^3
  +\mu_1^2+\mu_2^2+\mu_1\mu_2\right)
  +\mathcal{O}(\theta^3)\right\}\ .\nonumber
\end{eqnarray}

To test the above partition function by an independent computation,
we first take the scalars $X$ and $Y$ to be diagonal matrices
\begin{equation}
  X=\left(\begin{array}{cc}x&0\\0&-x\end{array}\right)\ ,\ \
  Y=\left(\begin{array}{cc}y&0\\0&-y\end{array}\right)\ .
\end{equation}
The relation is obviously satisfied. The gauge invariants are
\begin{equation}
  {\rm tr}(X^2)\sim x^2\ ,\ \ {\rm tr}(Y^2)\sim y^2\ ,\ \
  {\rm tr}(XY)\sim xy\ .
\end{equation}
A gauge-invariant operator made of these three units must have the
total number of $x$ and $y$ to be even, that is, either
\begin{equation}
  x^{2m+2}y^{2n+2}\ \ {\rm or}\ \ \ x^{2m+1}y^{2n+1}\ \ \ \
  (m,n=0,1,2\cdots)\ .
\end{equation}
Furthermore, for an operator with derivatives to be gauge-invariant,
the derivatives $\partial$ should effectively act on these units
only.

Let us check the sector with a single occurence of the derivative, that is, the
coefficient of $\theta$ in $Z_2(\mu_1,\mu_2,\theta)$. The following
set of operators
\begin{eqnarray}
  &&(\partial x)x^{2m+1}y^{2n}\ ,\ \
  (\partial x)x^{2m+2}y^{2n+1}\nonumber\\
  &&(\partial y)x^{2m}y^{2n+1}\ ,\ \
  (\partial y)x^{2m+1}y^{2n+2}\ \ \ \ (m,n=0,1,2\cdots)\nonumber
\end{eqnarray}
can be understood as containing the letter $\partial(x^2)$ or
$\partial(y^2)$, and containing even number of $x$'s and $y$'s. The
only other possible cases are those containing $\partial(xy)$, and
which cannot be written as one of the above four operators. They are
\begin{equation}
  \partial(xy)x^{2m}\ ,\ \ \partial(xy)y^{2n}\ \ \ \ (m,n=0,1,2\cdots;\
  (0,0)\ {\rm overcounted})
\end{equation}
since $(\partial x)y^{2n+1}$ and $(\partial y)x^{2m+1}$ terms in the
above cannot be rewritten to contain $\partial(x^2)$ or
$\partial(y^2)$ factors. The partition function for the above
operators, apart from the common $\theta$ factor, is
\begin{eqnarray}
  Z_2(\mu_1,\mu_2,\theta)\left.\frac{}{}\right|_{\theta^1}&=&
  \frac{\mu_2^2+\mu_2^2+\mu_1^3\mu_2+\mu_1\mu_2^3}{(1-\mu_1^2)(1-\mu_2^2)}+
  \frac{\mu_1\mu_2}{1-\mu_1^2}+\frac{\mu_1\mu_2}{1-\mu_2^2}-\mu_1\mu_2\nonumber\\
  &=&\frac{(1+\mu_1\mu_2)(\mu_1^2+\mu_2^2+\mu_1\mu_2-\mu_1^2\mu_2^2)}
  {(1-\mu_1^2)(1-\mu_2^2)}\ .
\end{eqnarray}
This agrees with the coefficient of $\theta$ in the proposed
partition function.

There is another way to derive (\ref{b1}) based on the Koszul complex 
(see for example \cite{Henneaux:1992ig}). One introduces a local 
{\it anticommuting} operator $C$ and a differential $\Delta$ such that 
\begin{eqnarray}
\Delta C &=&   {\rm tr}\left(X^2\right){\rm tr}\left(Y^2\right) - \left({\rm tr}(XY)\frac{}{}\right)^2\ , \nonumber \\
\Delta  {\rm tr}\left(X^2\right) &=&0\,, ~~~~
\Delta  {\rm tr}\left(Y^2\right) =0\,, ~~~~ 
\Delta  {\rm tr}(XY) =0\,.
 \end{eqnarray}
The operators $ {\rm tr}\left(X^2\right),  {\rm tr}\left(Y^2\right)$ and $ {\rm tr}\left(XY\right)$ 
are considered unrelated and the cohomology of the Koszul differential 
$H(\Delta)$ reproduces the space of constrained operators. The 
partition function $Z_2$ is now easily computed by taken into account 
the three free operators  $ {\rm tr}\left(X^2\right),  {\rm tr}\left(Y^2\right), {\rm tr}\left(XY\right)$, which are 
bosonic, and the fermionic operator $C$. The latter scales as $\mu_1^2 \mu_2^2$ if we scale $X$ and $Y$ with $\mu_1$ and $\mu_2$. So, combining these data, 
one obtains the formula (\ref{b1}): the three bosonic fields and their derivatives yield the 
contributions $(1- \mu_1^2 \theta^k)(1- \mu_2^2 \theta^k)(1- \mu_1 \mu_2 \theta^k)$ 
in the denominator of $Z_2$, while the anticommuting $C$ leads to the contribution 
$(1- \mu_1^2 \mu_2^2 \theta^k)$ in the numerator.

\section{Appendix to Section \ref{five}}
\label{D}
\subsection{Derivation of Equation (\ref{energythree})}
\label{partsappendix}

In this appendix, we derive equation (\ref{energythree}). The energy density (\ref{firstenergydensity}) can be rearranged as:
\begin{equation}\label{energycompletesquare}\begin{split}
  \mathcal{E}=&
  \frac{1}{2}\left(F_{12}+s_1\frac{1}{2}[\phi^i,\bar{\phi}^i]\right)^2
  -s_1{\rm tr}\left(\frac{}{} F_{12}[\phi^i,\bar{\phi}^i]\right)\\
  &+\frac{1}{2}\left(\frac{}{}F_{a0}-s_2 F_{a3}\right)^2
  +s_2 F_{a0}F_{a3}\\
  &+\frac{1}{2}\left|\frac{}{}D_0\phi^i-s_2 D_3 \phi^i+s_3 i\phi^i\right|^2+
  s_2\frac{1}{2}\left(D_0\phi^iD_3\bar{\phi}^i+D_0\bar{\phi}^iD_3\phi^i\right)\\
  &\hspace{1cm}-s_3\frac{i}{2}\left(\phi^iD_0\bar{\phi}^i-\bar{\phi}^iD_0\phi^i\right)
  +s_2s_3\frac{i}{2}\left(\phi^iD_3\bar{\phi}^i-\bar{\phi}^iD_3\phi^i\right)\\
  &+\frac{1}{2}\left|\frac{}{}(D_1+s_1iD_2)\phi^i\right|^2
  +is_1\left(D_1\phi^iD_2\bar{\phi}^i-D_2\phi^iD_1\bar{\phi}^i\right)\\
  &+\frac{1}{2}(F_{03})^2+\frac{1}{4}\left|\frac{}{}[\phi^i,\phi^j]\right|^2
\end{split}\end{equation}
where $s_{1,2,3}$ are $\pm$ signs and the trace on the gauge indices is understood.

We now note that
\begin{equation}\label{something}
  \int_{S^3}\tr D_1\phi^iD_2\bar{\phi}^i-\tr D_2\phi^iD_1\bar{\phi}^i
  =\int\tr\bar{\phi}^i[D_1,D_2]\phi^i-\epsilon_{ab}
  \partial_a{\rm tr}\left(\bar{\phi}^iD_b\phi^i\right)\ .
\end{equation}
The second term is
\begin{equation}\begin{split}
  \int_{S^3}\epsilon_{ab}E_a^\mu\partial_\mu{\rm tr}
  \left(E^\nu_b\bar{\phi}^i D_\nu \phi^i\right)&=
  \int\left(\det{e}\right)\left(\frac{}{}
  \epsilon_{ab}E_a^\mu(\partial_\mu E^\nu_b)
  {\rm tr}(\bar{\phi}^i D_\nu \phi^i)+\epsilon_{ab}
  E^\mu_a E^\nu_b\partial_\mu{\rm tr}
  \left(\bar{\phi}^iD_\nu \phi^i\right)\right)\\
  &=\int(\det{e})[\partial_1,\partial_2]^\nu{\rm tr}(\bar{\phi}^i
  D_\nu \phi^i)+\epsilon^{\mu\nu\rho}e^{3}_\rho\partial_\mu{\rm tr}
  \left(\bar{\phi}^iD_\nu \phi^i\right)\ ,
\end{split}\end{equation}
where $[\ ,\ ]^\nu$ denotes the Lie commutator of directional
derivatives and $E^\nu_b$ are components of the basis vectors of our local orthonormal frame
on the $S^3$. That is $E^\nu_b$ are the components of the dual vectors to the left-invariant
one forms listed in equation (\ref{leftinv}). Note that, up to a total derivative, the derivative
acting on $\tr\bar{\phi}^iD_\nu \phi^i$ in the last term can effectively be
regarded as acting on
\begin{equation}
  \tr\bar{\phi}^iD_\nu \phi^i\stackrel{eff}{=}\frac{1}{2}\left(
  \tr\bar{\phi}^iD_{\nu}\phi^i-\tr D_\nu\bar{\phi}^i \phi^i\right)\ .
\end{equation}
Combining these with the first term of equation (\ref{something}), which is
\begin{equation}
  \int\tr\bar{\phi}^i[D_1,D_2]\phi^i=\int -i{\rm \tr}(\bar{\phi}^i[F_{12},\phi^i])
  +[\partial_1,\partial_2]^\mu\tr\bar{\phi}^i D_\mu \phi^i\ ,
\end{equation}
one obtains
\begin{equation}\label{somethingtwo}
  \int_{S^3}{\rm tr}\Big(D_1\phi^iD_2\bar{\phi}^i-D_2\phi^iD_1\bar{\phi}^i\Big)=
  \int -i(\det{e}){\rm tr}\left(\frac{}{}\bar{\phi}^i[F_{12},\phi^i]\right)-
  \frac{1}{2}\epsilon^{\mu\nu\rho}e^{3}_\rho\partial_\mu{\rm tr}
  \left(\bar{\phi}^iD_\nu \phi^i-D_\nu\bar{\phi}^i \phi^i\right)\ .
\end{equation}
Note that the directional derivative
$\partial_3=\frac{\partial}{\partial\psi}$ generates an isometry. If
$\rho=\psi$, the second term is a total
derivative which can be ignored since $e^3_\psi=1$. If $\mu=\psi$,
it again becomes a total derivative because $e^3_\theta=0$
and $e^3_\phi$ is independent of $\psi$. Therefore, the only term which gives nontrivial
contribution is the term where $\nu=\psi$. So one can rearrange the second
term of equation (\ref{somethingtwo}) as
\begin{equation}
  +\int\frac{1}{2}\epsilon^{\mu\rho
  \psi}e^3_\rho\partial_\mu\left(\cdots\frac{}{}\right)_\nu=
  -\int\frac{1}{2}\epsilon^{\mu\rho
  \psi}(de^3)_{\mu\rho}\left(\cdots\frac{}{}\right)_\nu=
  +\frac{1}{2}\int(\det{e})\tr
  \left(\bar{\phi}^iD_\nu \phi^i-D_\nu\bar{\phi}^i \phi^i\frac{}{}\right)\ .
\end{equation}
Therefore, we finally obtain
\begin{equation}\label{somethingthree}
  i\int_{S^3}\tr D_1\phi^iD_2\bar{\phi}^i-\tr D_2\phi^iD_1\bar{\phi}^i=
  \int(\det{e}){\rm tr}\left(F_{12}[\phi^i,\bar{\phi}^i]+
  \frac{i}{2}\left(\frac{}{}
  \bar{\phi}^iD_\nu \phi^i-D_\nu\bar{\phi}^i \phi^i\right)\right)\ .
\end{equation}
Inserting this result into the 5th line of the complete-squared
energy functional (\ref{energycompletesquare}), one finds that the first term is canceled against
the same term in the first line of (\ref{energycompletesquare}). Furthermore, if $s_1=+s_2s_3$, the second term is
canceled with a cross term in the 4th line of (\ref{energycompletesquare}). The result is
\begin{eqnarray}
  \mathcal{E}&=&\tr\left[
  \frac{1}{2}\right.\left(F_{12}+s_1\frac{1}{2}[\phi^i,\bar{\phi}^i]\right)^2
  +\frac{1}{2}\left(\frac{}{}F_{a0}-s_2 F_{a3}\right)^2
  +\frac{1}{2}\left|\frac{}{}D_0\phi^i-s_2 D_3 \phi^i+s_3 i\phi^i\right|^2
  \nonumber\\
  &&\hspace{0.6cm}+\frac{1}{2}\left|\frac{}{}(D_1+s_1iD_2)\phi^i\right|^2
  +\frac{1}{2}(F_{03})^2+\frac{1}{4}\left|\frac{}{}[\phi^i,\phi^j]\right|^2\\
  &&\hspace{0.6cm}\left.
  +s_2\left( F_{a0}F_{a3}+\frac{1}{2}\left(D_0\phi^iD_3\bar{\phi}^i
  +D_0\bar{\phi}^iD_3\phi^i\right)\right)+s_3\frac{i}{2}\left(\bar{\phi}^iD_0
  \phi^i-\phi^iD_0\bar{\phi}^i\right)\right]\ .\nonumber
\end{eqnarray}
The signs $(s_1,s_2,s_3)$ are freely chosen as either $(+,+,+)$,
$(+,-,-)$, $(-,+,-)$ or $(-,-,+)$. Setting $(s_1,s_2,s_3)=(+,+,+)$ leads to
equation (\ref{energythree}). Other cases can be studied in a
similar way.

\subsection{Derivation of Gauss Constraint in $\mathbb{R}^4$}
\label{gaussapp}

We start from the equation of motion
\begin{equation}
 D^\mu F_{\mu\nu}+\frac{i}{2}\left([\phi^i,D_\nu\bar\phi^i]-
  [D_\nu\phi^i,\bar\phi^i]\frac{}{}\right)=0\ .
\end{equation}
Inserting $\nu=I$ and $\nu=\bar{I}$, one obtains
\begin{eqnarray}
  D_{J}F_{\bar{J}I}+\epsilon_{JI}D_{\bar{J}}F_{12}&=&
  -\frac{i}{4}\left([\phi^i,D_I\bar\phi^i]-
  [D_I\phi^i,\bar\phi^i]\frac{}{}\right)\\
  D_{\bar{J}}F_{J\bar{I}}&=&-\frac{i}{4}
  [\phi^i,D_{\bar{I}}\bar\phi^i]\ \label{secondgauss}.
\end{eqnarray}
Contracting $r^2Z^I$ with the first equation, imposing a BPS
equation and using the complex conjugation rule for field strengths,
one obtains
\begin{equation}
  \bar{\mathcal{D}}F_{12}-\mathcal{D}(F_{12})^\star=
  -\frac{i}{4}\left([\phi^i,Z\cdot D{\phi^i}^\star]-
  [Z\cdot D\phi^i,{\phi^i}^\star]-2[\phi^i,{\phi^i}^\star]\frac{}{}\right)
\end{equation}
where the curly derivatives denote
\begin{equation}
  \mathcal{D}\equiv\frac{1}{r^2}\left(\bar{Z}^2D_1-\bar{Z}^1D_2\right)
   \ ,\ \ \bar{\mathcal{D}}\equiv r^2
   \left(Z^2D_{\bar{1}}-Z^1D_{\bar{2}}\right)\ .
\end{equation}
This can be rewritten as
\begin{equation}\label{gaussonly}
  \left(\bar{\mathcal{D}}F_{12}+\frac{i}{4}[\phi^i,Z\cdot D{\phi^i}^\star]
  -\frac{i}{4}[\phi^i,{\phi^i}^\star]\right)-\left(\bar{\mathcal{D}}F_{12}
  +\frac{i}{4}[\phi^i,Z\cdot D{\phi^i}^\star]
  -\frac{i}{4}[\phi^i,{\phi^i}^\star]\right)^\star=0 \ .
\end{equation}
The second part of the Gauss constraint, (\ref{secondgauss}), can be
shown to be automatically satisfied by using the expression for
$F_{I\bar{J}}$ in terms of conjugate components, (\ref{fijconj}),
and the BPS equations $F_{I\bar{J}}\bar{Z}^J=0$ and
$D_{\bar{I}}\phi^i=0$. Therefore, all the equations of motion are
satisfied by solutions to the BPS equations except for the
constraint (\ref{gaussonly}) which comes from the Gauss law. This
constraint requires that the imaginary part of
$\bar{\mathcal{D}}F_{12}+\frac{i}{4}[\phi^i,Z\cdot
D{\phi^i}^\star]-\frac{i}{4}[\phi^i,{\phi^i}^\star]$ is zero. One can
actually show that this last quantity is imaginary, so that the
constraint from the Gauss law can simply be written as
\begin{equation}
  \bar{\mathcal{D}}F_{12}+\frac{i}{4}
  [\phi^i,Z\cdot D{\phi^i}^\star]-\frac{i}{4}[\phi^i,{\phi^i}^\star]=0\ .
\end{equation}

\section{Relation Between Spherical Coordinates and Complex Coordinates}
\label{coordapp}
The spherical coordinates, $(r,\theta,\psi,\phi)$, on $S^3\times \mathbb{R}$ used in section \ref{firstsection} are related to the complex coordinates of sections \ref{reform}, \ref{axialg}, \ref{classicalconfigs} and \ref{quantization} by:
\begin{equation}\begin{split}
Z^1&=r \cos\zeta\ e^{i{\psi+\phi\over 2}}\\
Z^2&=r \sin\zeta\ e^{i{\psi-\phi\over 2}}\ ,
\end{split}\end{equation}
where $\zeta=\theta/2$. This induces the following transformation of derivatives:
\begin{equation}\begin{split}
  \partial_\tau&=Z^I\partial_I+\bar{Z}^I\partial_{\bar{I}}\\
  \partial_3\equiv 2\partial_\psi&=i\left(
  Z^I\partial_I-\bar{Z}^I\partial_{\bar{I}}\right)\\
  \partial_1\equiv 2\partial_\theta&=
  -\left|\frac{Z^2}{Z^1}\right|\left(Z^1\partial_1+
  \bar{Z}^1\partial_{\bar{1}}\right)+
  \left|\frac{Z^1}{Z^2}\right|\left(Z^2\partial_2+
  \bar{Z}^2\partial_{\bar{2}}\right)\\
  \partial_2\equiv\frac{2}{\sin\theta}
  \left(\partial_\phi\!-\!\cos\theta\partial_\psi\right)&=
  -i\left|\frac{Z^2}{Z^1}\right|\left(Z^1\partial_1-
  \bar{Z}^1\partial_{\bar{1}}\right)+
  i\left|\frac{Z^1}{Z^2}\right|\left(Z^2\partial_2-
  \bar{Z}^2\partial_{\bar{2}}\right)\ ,
\end{split}\end{equation}
where $\partial_1,\partial_2,\partial_3$ are derivatives in the orthonormal frame (\ref{leftinv}) on $S^3$.
We also have
\begin{equation}
\bar{Z}\cdot\bar{\partial}={1\over 2}r\partial_r+i\partial_\psi\ ,
\end{equation}
which we use in section \ref{gaugeonly}.

\section{A Class of Exact $SU(2)_L$ Invariant Solutions}
\label{exactsolutions}

Now we will identify a set of exact BPS solutions to the
equations (\ref{BPSeqns}) and (\ref{gausseqn}) by
imposing additional symmetry.
We set the scalars to zero and consider solutions which preserve an
$SU(2)_L$ $\subset SO(4)$ symmetry,
rotating $Z^1$ and $Z^2$ as doublets. Since $A_I$ and $A_{\bar{I}}$
should transform as fundamental and anti-fundamental
representations, respectively, the only surviving terms in the
series expansions are
\begin{eqnarray}
  A_I&=&iM\frac{\bar{Z}^I}{r^2}+N\epsilon_{IJ}Z^J\label{sutwoone}\\
  A_{\bar{I}}&=&N^\ast\epsilon_{\bar{I}\bar{J}}
  \frac{\bar{Z}^J}{r^4}\ ,\label{sutwotwo}
\end{eqnarray}
where $M,N$ are constant matrices with no coordinate dependence
and $M$ is required to be a Hermitian matrix.

With the scalars set to zero, the Gauss' constraint on $F_{12}$
becomes $D_{\bar{I}}F_{12}=0$ and substituting (\ref{sutwoone})
and (\ref{sutwotwo}) we find
\begin{equation}\label{holomo}
  D_{\bar{I}}F_{12}=0\ \rightarrow\ \
  \left[N^\ast,2N+[M,N]\frac{}{}\right]=0\ ,
\end{equation}
where we used
\begin{equation}
  F_{12}=-2N-[M,N]\ \ .
\end{equation}
Now using the expression
\begin{equation}
  F_{I\bar{J}}=\left([M,N^\ast]-2N^\ast\right)
  \frac{\bar{Z}^I(\epsilon_{\bar{J}\bar{K}})\bar{Z}^K}{r^6}-iM\left(
  \frac{\delta_{I\bar{J}}}{r^2}-\frac{\bar{Z}^IZ^J}{r^4}\right)
  -i[N,N^\ast]\frac{(\epsilon_{IK}Z^K)
  (\epsilon_{\bar{J}\bar{L}}\bar{Z}^L)}{r^4}
\end{equation}
one finds that the condition $F_{1\bar{1}}+F_{2\bar{2}}=0$ becomes
\begin{equation}
  0=M+[N,N^\ast]\ .
\end{equation}
Inserting this back into the condition (\ref{holomo}) and rearranging, one
finds the final algebraic conditions on the (numerical)
matrices:
\begin{eqnarray}
  0&=&M+[N,N^\ast]\\
  4M&=&[N^\ast,[N,M]]+[N,[N^\ast,M]]\ ,
\end{eqnarray}
where we used $[N^\ast,[N,M]]=[N,[N^\ast,M]]$ at the second
equation, which can be derived from Jacobi identity and the first
equation. If we identify the matrices as
\begin{equation}
  M\sim Z^3\ ,\ \ N\sim X^1+iX^2\ ,
\end{equation}
where $X^a$ ($a=1,2,3$) are the three scalars in a BMN matrix model, these two
equations are exactly the same as the ones
obtained by solving the $1/16$ BPS equations of BMN matrix model in \cite{Bak:2005jh}.

We can count the degrees of freedom carried by this solution
by comparing the matrix variables and the number of equations. We
have $3N^2$ real numbers which should satisfy $2N^2$ equations ($2$
matrix equations), so we are left with $N^2$ real degrees of freedom.
However, since we can use the $U(N)$ global gauge transformation to
diagonalize one of the three matrices, we have $N$ gauge-invariant
degrees left. For instance, we may take advantage of this gauge
transformation to make $M$ diagonal.

There is a curious branch of additional modes if this diagonal
matrix $M$ vanishes. Then the counting the number of equations (that
is $2N^2$) in the previous paragraph does not work, since the second
equation simply disappears for $M=0$. In this case, the solution is a
diagonal matrix $N$, which is the $SU(2)_L$ invariant part of the
$U(1)^N$ solutions of the zero coupling limit of the gauge theory.

Collecting both branches of $SU(2)_L$ invariant solutions, we have found $2N$ modes,
some diagonal and others non-diagonal.

\section{Appendix to Section \ref{classicalconfigs}}
\label{G}
\subsection{Arguments Establishing the Obstruction at $\mathcal{O}(g_{YM})$}
\label{holomorphyconstraints}

Dividing the problem into two steps, we may rewrite equation
(\ref{linearorder}) as:
\begin{eqnarray}\label{heqone}
  h&=&\bar\delta f_1\ \ \ (\rightarrow\ h^\star=\delta f_1^\star)\\
  \delta\bar\delta h-\bar\delta\delta h^\star&=&
  i\left[\bar\delta f_0^\star,\delta f_0\right]+
  2i\left[f_0^\star,f_0\right]\ .\label{heqtwo}
\end{eqnarray}
The equation (\ref{heqtwo}) is a cousin of the `Laplace' equation
with source. We will state an integrability condition for this
equation, which restricts the right hand side of (\ref{heqtwo}).
Since $h$ has antiholomorphic degree zero, one finds from
(\ref{pseudolaplacian}) that $\delta\bar\delta$ acts on $h$ as a
Laplacian operator
$r^2\partial_I\partial_{\bar{I}}=\frac{1}{r}\partial_r r^3
\partial_r+\nabla_{S^3}$ which is diagonalized by the functions
\begin{equation}
\psi^m_{s,j_1,j_2}=r^m S^{j_1,j_2}_s\ ,
\end{equation}
where $S^{j_1,j_2}_s$ are the scalar spherical harmonics of rank $s$
in $\mathbb{R}^4$. We expand $h$ in terms of these functions and impose
the following requirements:
\begin{enumerate}
\item[(A)] $h$ must be degree zero in $\bar{Z}^I$.
\item[(B)] $h$ must belong to the image of the operator $\bar\delta$ since $h=\bar\delta f_1$.
\end{enumerate}
With these conditions, we would like to see which spherical
harmonics can appear on the left hand side of (\ref{heqtwo}).
The answer to this question will constrain the form of the
zeroth order solution $f_0(Z)$.

We first examine the condition (A). We choose the convention in
which the coordinate $\psi$ used in section \ref{firstsection}
corresponds to the angular momentum $j_1$ appearing in the spherical
harmonics $S^{j_1,j_2}_s$ (the coordinate relations may be found in
appendix \ref{coordapp}). This means that restricting to functions
of antiholomorphic degree zero requires
\begin{equation}\label{anglap}
\bar{Z}\cdot\bar{\partial}\psi^m_{s,j_1,j_2}=({1\over
2}r\partial_r+i\partial_\psi)r^m S^{j_1,j_2}_s =(m/2-j_1)r^m
S^{j_1,j_2}_s=0.
\end{equation}
Therefore a spherical harmonic expansion of $h$ will only contain
terms of the form $r^{2j_1}S^{j_1,j_2}_s$.

Now we study the condition (B). The form $h=\bar\delta
f_1=r^2\epsilon^{\bar{I}\bar{J}}Z^J\partial_{\bar{I}}f_1$ implies
that all terms in $h$ must contain at least 1 factor of either $Z^1$
or $Z^2$. This means that, in the Cartesian coordinate, terms of the
form $\left({\bar{Z}^1\over r^2}\right)^c\left({\bar{Z}^2\over
r^2}\right)^d$ do not appear in $h$. All other terms can appear in
$h$.

With $h$ satisfying the conditions (A) and (B), we want to know what kind
terms in the spherical harmonic expansion appear on the
left hand side of (\ref{heqtwo}). Since the eigenvalue of the
Laplacian operating on $\psi^m_{s,j_1,j_2}=r^m S^{j_1,j_2}_s$ is
$m(m+2)-s(s+2)$, the modes in $h$ which have $m=s$ or $m=-(s+2)$ are
annihilated by $\delta\bar\delta$ and do not appear in
$\delta\bar\delta h$. Now we see the following:

\begin{enumerate}
\item  From condition (A), $m=2j_1$, so harmonics with
$m=-(s+2)$ cannot appear in $h$ at all since $|2j_1|\leq s$. We
therefore conclude that it is the harmonics with $m=s$ that are
constrained not to appear in
$\delta\bar\delta h$. In Cartesian coordinates, these modes
$\psi^s_{s,j_1,j_2}=r^s S^{s/2,j_2}_s$ in $h$ correspond to purely
holomorphic terms, $(Z^1)^a(Z^2)^b$.
\item From condition (B), the terms $\left({\bar{Z}^1\over r^2}\right)^c\left({\bar{Z}^2\over
r^2}\right)^d$ do not appear in $h$ and therefore they cannot
appear in $\delta\bar\delta h$.
\end{enumerate}

All other $\psi^m_{s,j_1,j_2}$ which appear in a spherical harmonic expansion of $h$ are not
annihilated by the Laplacian $r^2\partial_I\partial_{\bar{I}}$ and
may appear in $\delta\bar\delta h$. In summary, all
terms $(Z^1)^a(Z^2)^b\left({\bar{Z}^1\over
r^2}\right)^c\left({\bar{Z}^2\over r^2}\right)^d$ may appear in
$\delta\bar\delta h$ except purely holomorphic terms ($c=d=0$),
and their $\star$ conjugates ($a=b=0$). Since $\bar\delta\delta h^\star=(\delta\bar\delta
h)^\star$, the same constraints apply to the term $\bar\delta\delta
h$ on the left hand side of equation (\ref{heqtwo}).

\subsection{Integrability Constraint without Scalars}
\label{integoneapp}
In this appendix, we explicitly extract the holomorphic portions of the right hand
side of equation (\ref{heqtwo}). We use the notation of section \ref{gaugeonly} and
first note a useful integral:
\begin{eqnarray}
  &&\int_{r=1}d^3\Omega
  Y_{k_1+n_1,k_2+n_2}(Y_{k_1,k_2})^\star(Y_{n_1,n_2})^\star\\
  &&=\sqrt{\frac{(n_1\!+\!n_2\!+\!1)!(k_1\!+\!k_2\!+\!1)!
  (n_1\!+\!n_2\!+\!k_1\!+\!k_2\!+\!1)!}
  {(2\pi^2)^3n_1!n_2!k_1!k_2!(n_1+k_1)!(n_2+k_2)!}}
  \int_{r=1}d^3\Omega|Z^1|^{2(n_1+k_1)}|Z^2|^{2(n_2+k_2)}\nonumber\\
  &&=\sqrt{\frac{1}{2\pi^2}\cdot\frac{(n_1\!+\!n_2\!+\!1)!
  (k_1\!+\!k_2\!+\!1)!}{n_1!n_2!k_1!k_2!}\cdot\frac{(n_1+k_1)!(n_2+k_2)!}
  {(n_1\!+\!n_2\!+\!k_1\!+\!k_2\!+\!1)!}}=
  \frac{C_{n_1n_2}C_{k_1k_2}}{C_{n_1\!+\!k_1,n_2\!+\!k_2}}
  \equiv c^{n_1n_2}_{k_1k_2}\ \ .\nonumber
\end{eqnarray}
This integral allows us to extract the holomorphic component of $2[f_0^\star,f_0]$, which is
\begin{equation}
  2[f_0^\star,f_0]\rightarrow\sum_{n_1,n_2=0}^\infty Y_{n_1n_2}(Z^1,Z^2)
  \sum_{k_2,k_2=0}^\infty 2c^{n_1n_2}_{k_1k_2}\left[a_{k_1k_2}^\ast,
  \frac{}{}a_{n_1\!+\!k_1,n_2\!+\!k_2}\right]\ .
\end{equation}
We can extract the holomorphic terms of $\left[\frac{}{}\bar\delta
f_0^\star,\delta f_0\right]$ in a similar manner. Acting the operator
$\delta$ on $Y_{n_1n_2}$, one obtains
\begin{equation}
  \delta Y_{k_1k_2}=\sqrt{k_1(k_1\!+\!k_2\!+\!1)}\
  \frac{\bar{Z}^2}{r^2}\ Y_{k_1\!-\!1,k_2}-\sqrt{k_2(k_1\!+\!k_2\!+\!1)}\
  \frac{\bar{Z}^1}{r^2}\ Y_{k_1,k_2\!-\!1}\ .
\end{equation}
The coefficients of the holomorphic term proportional to
$Y_{n_1n_2}$ can be obtained by observing that the following
integral gives the only nonzero contribution:
\begin{eqnarray}
  &&\hspace{-1cm}\int_{r=1}d^3\Omega
  \left(\delta Y_{k_1+n_1,k_2+n_2}\right)(\delta Y_{k_1,k_2})^\star
  (Y_{n_1,n_2})^\star\nonumber\\
  &&\hspace{-0.5cm}=C_{n_1n_2}C_{k_1k_2}C_{n_1\!+\!k_1,n_2\!+\!k_2}\int_{r=1}
  d^3\Omega\left(\frac{}{}k_1(k_1\!+\!n_1)
  |Z^1|^{2(n_1+k_1-1)}|Z^2|^{2(n_2+k_2+1)}\right.\nonumber\\
  &&\left.+k_2(k_2\!+\!n_2)
  |Z^1|^{2(n_1+k_1+1)}|Z^2|^{2(n_2+k_2-1)}
  -\left(k_1(k_2\!+\!n_2)+k_2(k_1\!+\!n_1)\frac{}{}\right)
  |Z^1|^{2(n_1+k_1)}|Z^2|^{2(n_2+k_2)}\frac{}{}\right)\nonumber\\
  &&\hspace{-0.5cm}=C_{n_1n_2}C_{k_1k_2}C_{n_1\!+\!k_1,n_2\!+\!k_2}
  \left(\frac{k_1(k_1\!+\!n_1)}{(C_{n_1\!+\!k_1\!-\!1,n_2\!+\!k_2\!+\!1})^2}+
  \frac{k_2(k_2\!+\!n_2)}{(C_{n_1\!+\!k_1\!+\!1,n_2\!+\!k_2\!-\!1})^2}-
  \frac{k_1(k_2\!+\!n_2)+k_2(k_1\!+\!n_1)}
  {(C_{n_1\!+\!k_1,n_2\!+\!k_2})^2}\right)\nonumber\\
  &&\hspace{-0.5cm}=c^{n_1n_2}_{k_1k_2}\left(
  \frac{}{}k_1(n_2\!+\!k_2\!+\!1)+k_2(n_1\!+\!k_1\!+\!1)-
  \left(k_1(k_2\!+\!n_2)+k_2(k_1\!+\!n_1)\frac{}{}\right)\right)
  =(k_1\!+\!k_2)c^{n_1n_2}_{k_1k_2}\ .
\end{eqnarray}
Note that the above coefficient is just zero for the case
$(k_1,k_2)=(0,0)$. With this integral in hand, we collect all the
holomorphic parts of $\left[\frac{}{}\bar\delta
f_0^\star,\delta f_0\right]+2[f_0^\star,f_0]$,
\begin{eqnarray}
  &&\hspace{-1cm}\left[\bar\delta f_0^\star,\delta
  f_0\right]+2[f_0^\star,f_0]\\
  &&\hspace{-1cm}\rightarrow\sum_{n_1,n_2=0}^\infty
  Y_{n_1n_2}(Z^1,Z^2)\sum_{k_1,k_2=0}^\infty
  \left(k_1\!+\!k_2\!+\!2\right)c^{n_1n_2}_{k_1k_2}
  \left[a_{k_1k_2}^\ast,\frac{}{}a_{n_1\!+\!k_1,n_2\!+\!k_2}\right]
  \nonumber\\
  &&\hspace{-1cm}=\sum_{n_1,n_2=0}^\infty Y_{n_1n_2}(Z^1,Z^2)
  \sum_{k_1,k_2=0}^\infty\left(k_1\!+\!k_2\!+\!2\right)
  \frac{C_{n_1n_2}C_{k_1k_2}}{C_{n_1\!+\!k_1,n_2\!+\!k_2}}
  \left[a_{k_1k_2}^\ast,\frac{}{}a_{n_1\!+\!k_1,n_2\!+\!k_2}\right]
  \ ,\nonumber
\end{eqnarray}
which is the result that appears in equation (\ref{gaugeonlyconstraint}).

\subsection{Integrability Constraint with Scalars}
\label{scalarapp}
In this appendix, we extract the purely
holomorphic terms from the source appearing on the right hand side of
equation (\ref{perttwosecond}) in the most general bosonic case, where
all scalars and the gauge field may be nonzero. The treatment of the
$[{\phi^i_0}^\star,\phi^i_0]+[\bar\delta{\phi^i_0}^\star,\delta\phi^i_0]$
part of the source is very similar to that of appendix \ref{integoneapp}.
It gives the following holomorphic terms
\begin{eqnarray}
  &&\hspace{-1.5cm}\left[\bar\delta f_0^\star,\delta
  f_0\right]+2[f_0^\star,f_0]+\frac{1}{4}\left(
  [{\phi^i_0}^\star,\phi^i_0]+
  [\bar\delta{\phi^i_0}^\star,\delta\phi^i_0]\frac{}{}\right)\\
  &&\hspace{-1.5cm}\rightarrow\sum_{n_1,n_2=0}^\infty
  Y_{n_1n_2}\sum_{k_1,k_2=0}^\infty
  c^{n_1n_2}_{k_1k_2}\left(
  \left(k_1\!+\!k_2\!+\!2\right)
  \left[a_{k_1k_2}^\ast,\frac{}{}a_{n_1\!+\!k_1,n_2\!+\!k_2}\right]
  +\frac{1}{4}\left(k_1\!+\!k_2\!+\!1\right)
  \left[b_{k_1k_2}^{i\ast},\frac{}{}b^i_{n_1\!+\!k_1,n_2\!+\!k_2}\right]\right)
  \nonumber
\end{eqnarray}
where we are using the same notation as in section \ref{gaugeonly}.
The remaining term in the source takes the form
$(\delta\bar\delta+\bar\delta\delta)(\cdots)$, where
\begin{equation}
  \delta\bar\delta+\bar\delta\delta\stackrel{eff}{=}
  2r^2\partial_I\partial_{\bar{I}}-(Z\cdot\partial)\ .
\end{equation}
We now collect the purely holomorphic portions
of this term.
\begin{equation}
  \frac{i}{4}(\delta\bar\delta+\bar\delta\delta)[\phi^i_0,
  {\phi^i_0}^\star]^\prime=\frac{i}{4}\left(
  \frac{}{}2r^2\partial_I\partial_{\bar{I}}-
  (Z\cdot\partial)\right)[\phi^i_0,{\phi^i_0}^\star]^\prime
  =F(Z)+F^\star(\frac{\bar{Z}}{r^2})\ ,
\end{equation}
where the prime denotes only keeping the terms in $[\phi,\phi^\star]$
which would give holomorphic plus conjugate terms in the right hand
side. We have seen in section \ref{gaugeonly} that $\partial_I\partial_{\bar{I}}$
cannot produce any such terms and therefore the holomorphic part of the
$\delta\bar\delta+\bar\delta\delta$ contribution to the source is simply
the holomorphic part of
\begin{equation}
  -\frac{i}{4}(Z\cdot\partial)[\phi^i_0,{\phi^i_0}^\star
  ]\ .
\end{equation}
Collecting all contributions, we find that the holomorphic part of the source is
\begin{eqnarray}
  &&\hspace{-1.5cm}\left[\bar\delta f_0^\star,\delta
  f_0\right]+2[f_0^\star,f_0]+\frac{1}{4}\left(
  [{\phi^i_0}^\star,\phi^i_0]+
  [\bar\delta{\phi^i_0}^\star,\delta\phi^i_0]\frac{}{}\right)
  +\frac{1}{4}(Z\cdot\partial)[{\phi^i_0}^\star,\phi^i_0
  ]\ .\\
  &&\hspace{-1.5cm}\rightarrow\sum_{n_1,n_2=0}^\infty
  Y_{n_1n_2}\sum_{k_1,k_2=0}^\infty \nonumber \\
  && \times \ c^{n_1n_2}_{k_1k_2}\left(
  \left(k_1\!+\!k_2\!+\!2\right)
  \left[a_{k_1k_2}^\ast,\frac{}{}a_{n_1\!+\!k_1,n_2\!+\!k_2}\right]
  +\frac{1}{4}\left(n_1\!+\!n_2\!+\!k_1\!+\!k_2\!+\!1\right)
  \left[b_{k_1k_2}^{i\ast},\frac{}{}b^i_{n_1\!+\!k_1,n_2\!+\!k_2}\right]\right)
  \nonumber\\
\end{eqnarray}


\end{document}